%% file: main.tex
\RequirePackage{fix-cm}
\documentclass[twocolumn,epjc3]{svjour3}  
\smartqed  
\RequirePackage{graphicx}
\usepackage{units}
\usepackage{siunitx}
\usepackage{booktabs}
\usepackage{subfig}
\usepackage{textcomp}
\usepackage{multirow}
\usepackage{placeins}

\usepackage[version=4]{mhchem}
 \RequirePackage{mathptmx}      
\journalname{Eur. Phys. J. C}
\begin{document}

\title{The Design and Sensitivity of JUNO's scintillator radiopurity pre-detector OSIRIS
}


\author{The JUNO Collaboration}
\institute{Juno\_pub\_comm@juno.ihep.ac.cn}




\date{Received: date / Accepted: date}

\maketitle

\begin{abstract}
The OSIRIS detector is a subsystem of the liquid scintillator fillling chain of the JUNO reactor neutrino experiment. Its purpose is to validate the radiopurity of the scintillator to assure that all components of the JUNO scintillator system work to specifications and only neutrino-grade scintillator is filled into the JUNO Central Detector. The aspired sensitivity level of $10^{-16}$\ g/g of $^{238}$U and $^{232}$Th requires a large ($\sim$20\,m$^3$) detection volume and ultralow background levels. The present paper reports on the design and major components of the OSIRIS detector, the detector simulation as well as the measuring strategies foreseen and the sensitivity levels to U/Th that can be reached in this setup.

\end{abstract}

\begin{figure*}[h!]
    \centering
    \includegraphics[width=\textwidth]{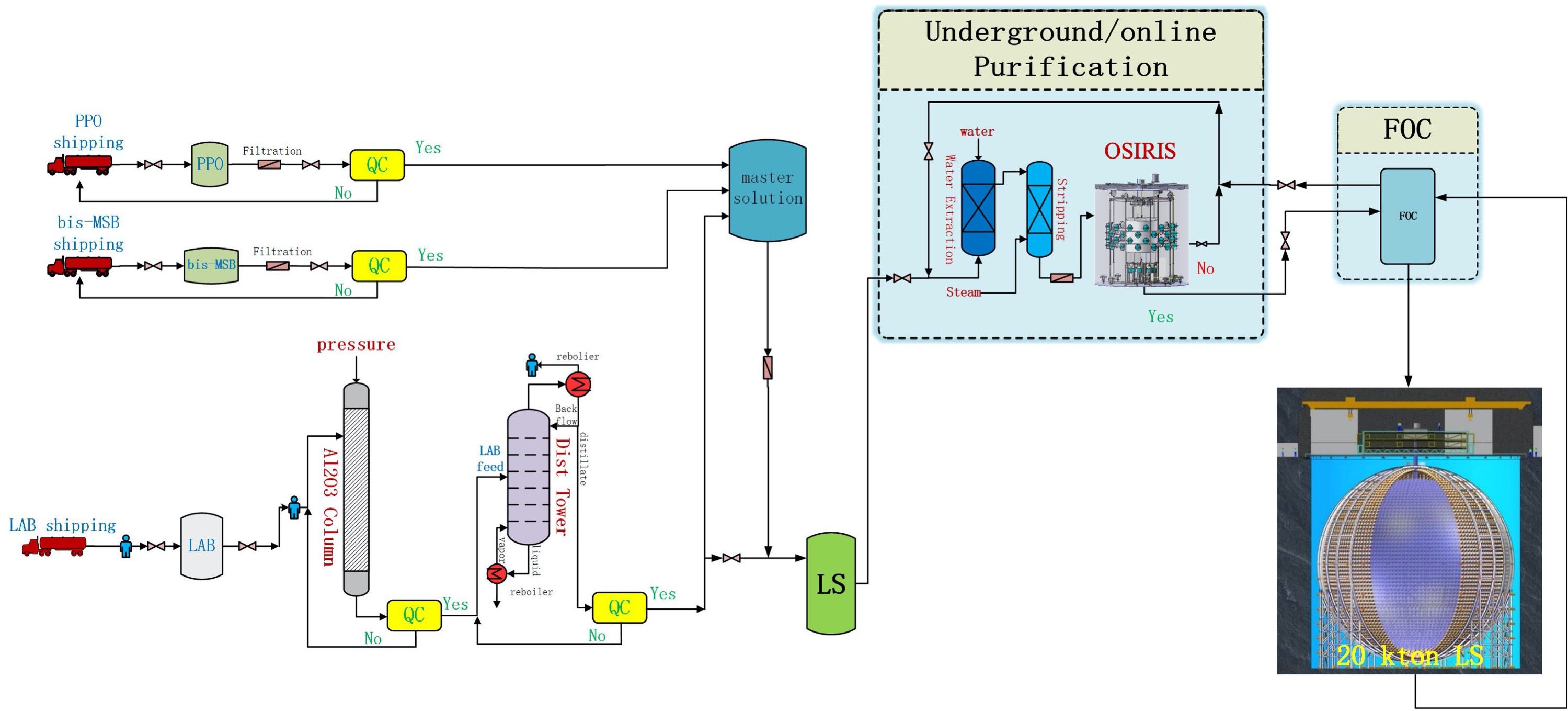}
    \caption{Layout of the JUNO scintillator system including four purification stages (Al$_2$O$_3$ column, distillation, water extraction, steam stripping) as well as storage and mixing tanks. OSIRIS is located as last stage in the filling line to monitor the product radiopurity. From there, the LS will be passed on to the JUNO filling system (FOC)}
    \label{fig:LS_layout}
\end{figure*}

\section{Introduction}
\label{sec:introduction}
The Jiangmen Underground Neutrino Observatory (JUNO) \cite{An:2015jdp} is a dedicated experiment to measure the neutrino mass hierarchy. JUNO will observe electron antineutrinos emitted by several nuclear reactor cores at a distance of $\sim$\unit[53]{km} with a 20 kiloton liquid scintillator (LS) detector. In such a measurement shielding against external radiation and careful selection of radiopure detector materials play a crucial role. Radioactivity in LS may result in events mimicking the inverse beta decay (IBD) coincidence signals of electron antineutrinos or cause pile-up to the single events distorting the energy scale, both having direct impact on the sensitivity of the neutrino mass hierarchy measurement. Requirements are even more demanding for other physics objectives, especially solar neutrino detection. This is why stringent upper limits have been set on the contamination level of the LS with uranium and thorium chain elements: $\leq$10$^{-15}$\, g/g for the IBD-based physics program ({\it IBD-level}) and $\leq$10$^{-16\div17}$\, g/g for solar neutrino detection ({\it solar-level}) (Tab.~\ref{tab:radiopurity}).

To achieve this radiopurity requirements an extensive purification program needs to be performed on the LS before filling the JUNO detector. This requires to set up a chain of specialized purification plants at the JUNO experimental site: column chromatography in alumina, distillation, water extraction and steam stripping. As the last stage in this purification chain, the Online Scintillator Internal Radioactivity Investigation System (OSIRIS) will serve as a stand-alone detector to verify the efficiency of the upstream purification plants and to monitor the radiopurity of the product LS during the filling of the JUNO Central Detector. 

The OSIRIS setup has been optimized for the purpose to detect the residuals of natural U/Th contamination of the LS. It will hold a 18\,t sample of LS enclosed in an Acrylic Vessel (AV) for screening, surrounded by an extensive water shield. The search is based on the fast coincidence decays of $^{214}$Bi-$^{214}$Po and $^{212}$Bi-$^{212}$Po present in the decay chains of U and Th, respectively. The design of OSIRIS permits two operation modes: {\it batch mode} and {\it continuous mode}. In {\it batch mode}, a single 18-t sample of LS will be monitored over several days or weeks. It is the preferred mode of operation for the commissioning phase of the JUNO purification chain since it allows a precise determination of scintillator quality and purification efficiency, reaching down to {\it solar-level} radiopurity. Instead, the {\it continuous mode} will be implemented during the JUNO filling phase since the constant exchange of LS in the AV will permit uninterrupted monitoring and warning capability in case of a malfunction of the purification systems. The corresponding sensitivity is reduced compared to the {\it batch mode} but will approach {\it IBD-level}.

The purpose of the present paper is to provide an over-view both of the design and the resulting sensitivity of the OSIRIS detector setup. We first provide an overview of the  main components of the OSIRIS detector setup (Sec.~\ref{sec:design}) before we go into a more detailed description of the experimental subsystems and foreseen operation modes of OSIRIS (Sec.~\ref{sec:components}). We present as well the detailed Geant4-based detector simulation that has been used to optimize the detector design and will serve for the preparation of the OSIRIS analysis (Sec.~\ref{sec:simulation}). Finally, the resulting sensitivities for U/Th screening as well as detecting contaminations from $^{14}$C and $^{210}$Po are reported in section \ref{sec:sensitivity}.

\section{Design Overview}
\label{sec:design}

The OSIRIS setup forms the last stage of the JUNO purification chain (fig.~\ref{fig:LS_layout}), monitoring the final product of the scintillator mixing and purification plants that are mounted on-site both on surface and below ground. The detector will be located in the rear of the Liquid Scintillator Hall of the JUNO underground laboratory, i.e.~in close proximity to the Water Extraction and Steam Stripping purification plants. An isometric view as well as the Scintillator Hall layout plan can be found in Appendix \ref{app:LS_hall}. Given the underground conditions, footprint and height of the OSIRIS facility are limited. On the other hand, the detector will benefit from about 700\,m of rock shielding, a crucial pre-requisite to reach sufficiently low background levels for its measurement program.

\begin{figure}[t]
    \centering
    \includegraphics[width=0.45\textwidth]{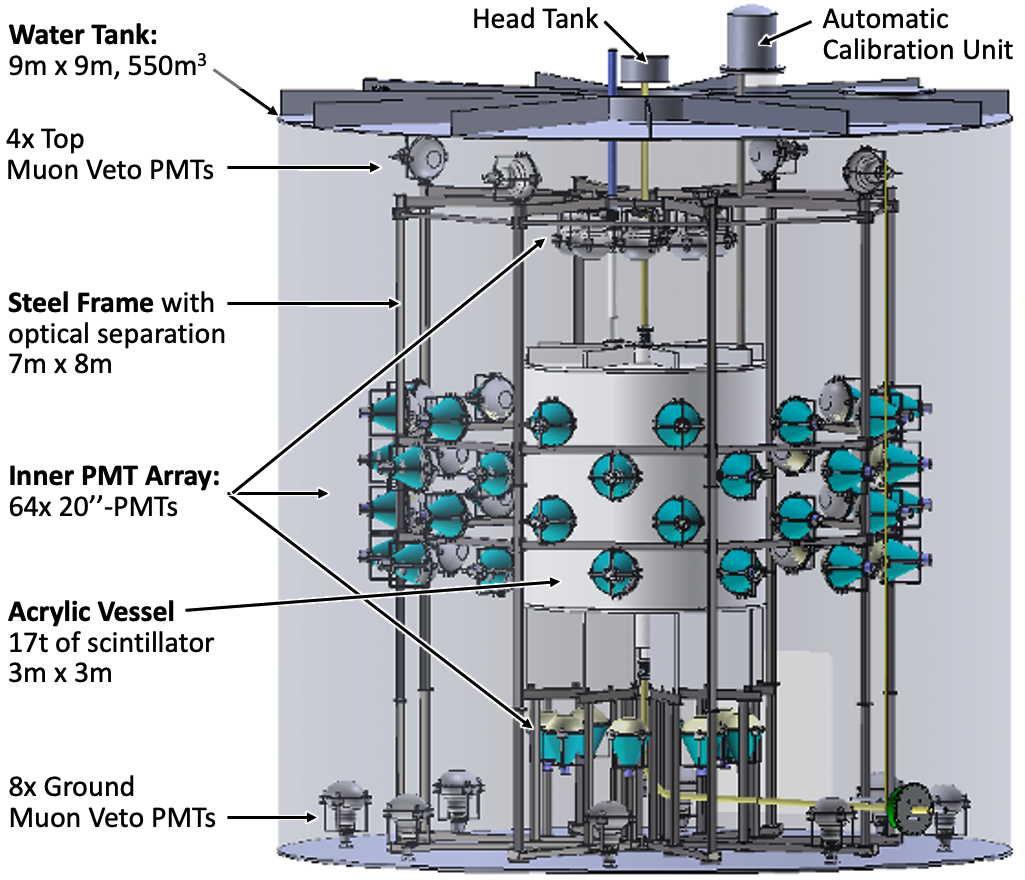}\hspace{2.2cm}
    \caption{Layout of the OSIRIS detector. The detector is segregated into the Inner Detector (ID) that is formed by the LS volume and the inner PMT array and the Outer Detector (OD) that uses the shielding water as a Cherenkov muon veto}
    \label{fig:osiris_layout}
\end{figure}

The detector dimensions are outlined by the outer Water Tank that features about 9.4\,m diameter and height. A simplified depiction of the interior is shown in figure \ref{fig:osiris_layout}. The setup is logically divided into an Inner Detector (ID) containing the LS volume and surrounding PMTs and an Outer Detector (OD) equipped with few PMTs and utilizing the water shielding as a Cherenkov muon veto. They are physically delimited by the Steel Frame and the attached Optical Separation. 

\paragraph{Scintillator target} Similarly to the Borexino Counting Test Facility (CTF) design \cite{Alimonti:1998aa}, the LS batch to be tested is contained in a transparent vessel with an inner volume of 21\,m$^3$ (18 tons of LS), located in the center of the setup. Given the considerably larger LS volume, counting statistics are accumulated faster than in the case of CTF (4.8\,m$^3$), allowing for a shorter average measuring time. This inner counting volume is surrounded by a water buffer for shielding, separated from the LS by the Acrylic Vessel.

\paragraph{Acrylic vessel (AV)} The cylindrical vessel of 3\,m height and diameter holds the LS at the center of the detector. An acry-lic wall thickness of 3\,cm and external stiffeners provide sufficient sturdiness for the frequent filling and LS exchange operations. The vessel is held in place by eight vertical acry-lic plates of 1\,m height, which are in turn mounted on the inner Steel Frame of 2\,m height (Sec.~\ref{sec:av}).

\paragraph{Steel Frame (SF)} The octagonal frame consists of rectangular stainless steel profiles. It reaches a height of 8\,m and a diameter of 7\,m (Sec.~\ref{sec:mechanics}). While its inner section supports the AV, the outer section holds the photomultipliers, calibration and sensor systems as well as the black-and-white PET sheets constituting the Optical Separation between ID and OD.  

\paragraph{Inner PMT Array} The scintillation light produced by events inside the LS volume is recorded by 64 20''-Hamamatsu PMTs mounted to the SF and facing inwards. For shielding the $\gamma$-rays emitted from PMT glass, the tubes are placed at a distance of 1.3\,m from the AV surface (i.e.~roughly 2.8\,m from the detector center). This configuration corresponds to a photoactive coverage of $\sim$9\,\% of the full solid angle, translating to a photo electron (p.e.) yield of $\sim$280\ p.e./MeV. OSIRIS will use a novel intelligent PMT design that permits digitization of the signals directly at the PMT base (Sec.~\ref{sec:pmts}).

\paragraph{Water Tank (WT)} The cylindrical WT holds a volume 9\,m in height and diameter, offering a $4\pi$ shielding of $\geq$3\,m of water from external gamma rays emitted by the cavern rock of the Scintillator Hall. The residual $\gamma$-flux has been determined to be sufficiently low for an effective Bi-Po coincidence search (Sec.~\ref{sec:simulation} and \ref{sec:sensitivity}). The tank is made from bolted carbon steel plates. They are protected from contact with the ultrapure water by a \unit[3]{mm} thick HDPE liner to prevent corrosion. The tank features several flanges and feed-throughs for pipes and cabling (Sec.~\ref{sec:mechanics}).

\paragraph{Muon Veto} Cosmic muons, secondary neutrons and radioactive isotopes created in spallation processes feature a finite probably to mimic the fast coincidence signals of the Bi-Po decays. Thus, a secondary array of 12 20''-PMTs watches the volume between SF and tank walls, using the ultrapure water as a Cherenkov radiator for crossing muons. To enhance light collection a layer of Tyvek is fastened to the WT walls. Moreover, the white outer surfaces of the Optical Separation enhance light reflection (Sec.~\ref{sec:muonveto}) .

\paragraph{Calibration systems} In order to interpret the acquired data, picosecond laser pulses will be inserted into the detector by a system of optical fibers with the emission points mounted to the SF. Moreover, an Automatic Calibration Unit (ACU)\cite{LIU201419} refurbished from the Daya Bay experiment is connected to the AV via a steel pipe in order to lower radioactive sources and an LED directly into the LS volume to calibrate the detector response.  Source positions can be cross-checked using a CCD system mounted to the SF (Sec.~\ref{sec:calibration}).

\paragraph{Liquid Handling System (LHS)} The relative fragility of the AV and different operation modes envisaged for the JUNO commissioning and filling phases requires a sophisticated system to handle the LS and water in AV and tank. In order to support the formation of temperature stratification levels of the LS in the AV, a constant flow of warm LS inserted to the top of the AV is countered by an upflow of cooler water on the outside of the AV. A head tank is located above the AV to equalize hydrostatic pressure between AV LS and WT water. A gas handling system provides buffers of ultrapure nitrogen gas to both volumes at a slight overpressure to prevent contamination of the liquids with environmental radon (Sec.~\ref{sec:lhs}).

\section{Detector Components}
\label{sec:components}

\subsection{Mechanical design}
\label{sec:mechanics}

The basic dimensions of the OSIRIS facility are governed by the requirements of the Bi-Po coincidence search. Here, we describe the conceptual design of the basic mechanical components. The outer dimensions of the setup are given by the Water Tank (WT). The WT holds the water buffer required to shield the LS volume from external radioactivity and serves as a Cherenkov radiator for the Outer Detector (Sec.~\ref{sec:muonveto}). In its center, the LS batch under investigation will be contained in the transparent Acrylic Vessel (AV) described in Sec.\,\ref{sec:av}. The AV is held in place and surrounded by the Steel Frame (SF) that also serves as support for the PMTs and calibration systems discussed in later sections.

\subsubsection{Water Tank}
\label{sec:watertank}

\begin{figure*}[t!]
    \centering
    \includegraphics[width=\textwidth]{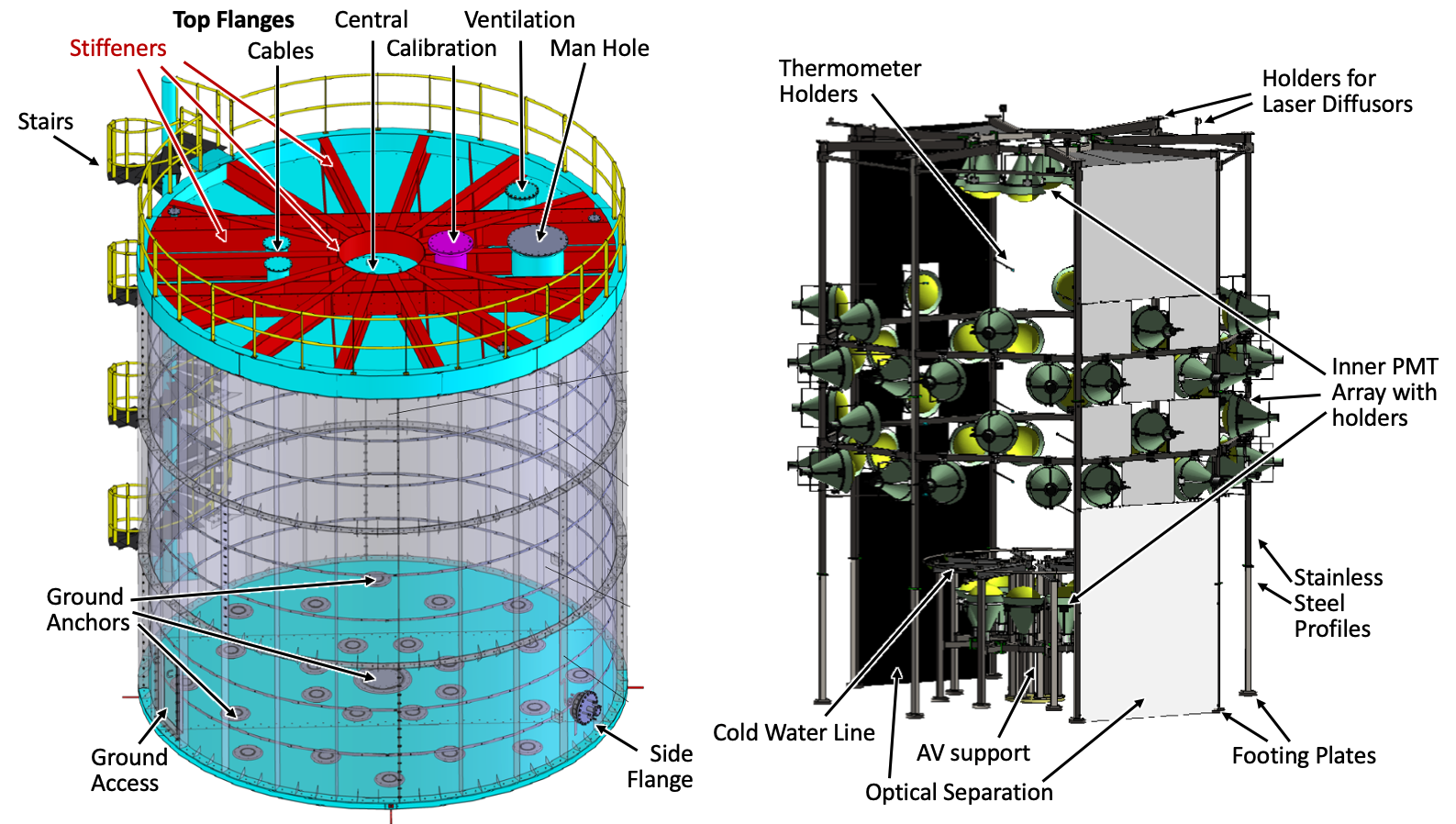}
    \caption{Main elements of the OSIRIS Water Tank and Steel Frame}
    \label{fig:tankandframe}
\end{figure*}

\paragraph{Overall dimensions} The outer water tank is assembled from flat and curved carbon steel elements (Q235A/SS400) that are connected by bolts. The parts have been already pre-fabricated by the chinese Jiujiang company. The tank consists of 3 bottom, 12 wall and 3 roof elements. The steel sheets themselves are 4--8\,mm in thickness, but perpendicular stiffeners are attached to all pieces in order to provide the required mechanical stability. Stiffeners running around the perimeter of the tank are 200\,mm height. The thickest stiffeners of 500\,mm in height are included at the top to bear the additional weight by the top installations (see below). So while the inner water volume is designed to be 9\,m height and diameter, the outer dimensions are about 9.6\,m in height and 9.4\,m in diameter. The overall layout is displayed in the left panel of Fig.~\ref{fig:tankandframe}.

\paragraph{Internal Liner} In order to prevent corrosion of the tank \, structure due to direct contact of the carbon steel with the ultrapure water, all inner tank surfaces are covered by a \unit[3]{mm} strong HDPE liner (Geotech Lining Corporation). Individual pieces are heat-welded together, providing both liquid and gas tightness (see below). Fastening of the flexible liner elements on the walls is achieved by steel battens. Special care has to be given at the interfaces: Flanges are covered by individually preformed HDPE elements, and the closing blind flanges are made from stainless steel. The basic color of the liner is black but highly reflective Tyvek sheets are heat-welded to most surfaces to increase light collection in the Outer Detector. 

\paragraph{Interfaces} The WT requires both inside interfaces for mechanical installations and feed-throughs to the surroundings. Internally, a total of 25 ground anchor plates permits the mounting of the SF and the ground muon veto PMTs. The interfaces offered to the inside water are stainless steel discs from 200 to 890\,mm diameter with thread patterns for moun-ting the SF and veto PMTs and circumferential seams to seal the HDPE liner for water-tightness. Smaller pieces (like thermometers) that are mounted on the tank walls will be held in place by welded-on HDPE holders. A selection of flanges in the tank roof provides interfaces to the outside: a 1\,m-diameter Central Flange supports the Head Tank and piping for filling of the AV; the neighbouring Calibration Flan-ge of 70\,cm diameter holds the ACU; two 40\,cm diameter flan-ges provide the feed-throughs for all internal cables by crimp connectors; a 50\,cm flange is used for tank ventilation during the construction phase. One inlet and three outlet pipes on the top of the tank permit to establish a nitrogen blanket above the water. Access to the tank during operation is possible via a 1\,m-diameter Man Hole. During construction, additional access is provided by a removable panel element (1\,m$\times$1.5\,m) at ground level. In addition, there is a Side Flange that is used both for the water operations in the outer tank volume and the draining of the AV. 

\paragraph{Top Installations} The WT roof has been strengthened to bear the weight of several on-top structures. The Top Clean Room (TCR) with a footprint of 5\,m$\times$3.5\,m protects the Central Flange, Calibration Flange, and Man Hole as well as holding several LS and gas systems that are part of the LHS (Sec.\ \ref{sec:lhs}). It is also equipped with a small pre-room (2.5\,m$\times$1.5\,m) for the access of personnel and equipment. Next to the TCR, an open platform of 5\,m$\times$3\,m area supported by aluminum grids will hold the electronics required for iPMT read-out as well as the detector DAQ and slow control (Secs.~\ref{sec:pmts}, \ref{sec:daq}). A spiral staircase has been added to the side of the WT for access to the tank roof during construction and operation. While heavier items required for the construction and installation phases are moved by a central hall crane, smaller items ($<$375\,kg) are lifted by a dedicated crane trolley running on a 3\,m-long I-beam suspended from the hall roof.

\paragraph{Other design requirements} The WT acts as the most important barrier to shield the LS in the AV from potential impurities of the surroundings. During the OSIRIS construction phase, an effective clean room of class 10,000 (ISO7) will be established in the interior. The gas tightness that is to be maintained during detector operation is on the level of $10^{-4}$\ mbar$\cdot$l/s in order to prevent external radon from migrating into the water buffer. This is achieved with Viton-A, Kalrez and PTFE O-rings in the flanges, crimp connectors for the cable feed-throughs and the HDPE liner that covers potential gaps between the steel elements. In addition, a slight overpressure of +5\,mbar is maintained in the nitrogen buffer relative to the surroundings.

In accordance to the filling specifications of the JUNO main detector, the interior of the tank is to be kept at a temperature of 21\,$^\circ$C. This is about 10\,$^\circ$C lower than the temperature of the rock on which the WT is resting. To maintain this temperature difference, two layers of passive heat insulation materials (flexible elastomeric foam and polyisocyanurate foam) are added below the tank. In addition, cool water is inserted through a star-shaped pipe system on the ground of the WT to compensate any heat permeating the insulation layers.

Additional shielding from external gamma rays has been added below the insulating layers. The shielding consists from three layers of carbon steel plates embedded in the concrete floor below the tank. The maximum strength below the detector center is 14\,cm, the dimension of the lower-most steel layer 5.5\,m$\times$5.5\,m. The additional material reduces the rock-induced external background entering from below by two orders of magnitude.

\subsubsection{Steel Frame}
\label{sec:steelframe}

The Steel Frame (SF) serves two primary functions. The inner part below the AV supports the vessel when both WT and AV are empty but also balances buoyancy forces in the filled configuration. It is an octagonal structure with an outer diameter of 3\,m and a height of 2\,m, placing the LS volume right at the center of the WT and providing support for eight upward-facing inner PMTs. The outer part supports most PMTs, the laser calibration system and other sensors, and serves as separation between Inner and Outer Detectors. The latter is an octagonal structure with a height of 8.1\,m, side length of 2.4\,m, and a maximum diameter of 5.8\,m. The layout of the overall structure is depicted in the right panel of Fig.~\ref{fig:tankandframe}. The eight vertical profile sections are joined together by two rings of horizontal profiles at medium height (to mount four rings of twelve PMTs each) as well as a star and ring structure at the top that supports the 8 downward-facing inner PMTs as well as the four top muon veto PMTs. The frame will be manufactured by Jiujiang company.

\paragraph{Steel profiles} Both subsections of the SF are assembled from stainless steel (SUS316L, 1.4404) hollow profiles of 80\,mm edge length, 4\,mm wall thickness, and up to 3.45\,m in bar length. Most are terminated on both ends by quadratic or rectangular plates with bore holes through which adjacent elements can be joined by M12 screws and nuts. The lowest elements provide larger footing plates to provide additional stability to the connection to the WT anchor plates. All surfaces are mechanically polished to a roughness of \SI{0.4}{\micro\meter} in order to limit radon emanation and to prevent corrosion in the ultrapure water. Moreover, the U/Th content of the steel has been specified to be below 10\, ppb.

\paragraph{Adapters} Most of the installations attached to the SF require additional holding or adapter elements that provide for correct mounting and orientation: the PMTs, the diffusers of the fiber calibration system, the inner thermometer array, and the circular pipe that is mounted at the upper edge of the inner SF to provide an outflow of cool water along the surface of the AV. All these adapter parts will be made from electro-polished stainless steel (SUS316L or SUS316Ti) and fastened to steel bolts welded directly onto the frame profiles. 

\paragraph{Optical Separation} The water volume is divided into two optically separated regimes by an array of thin PET sheets (\SI{200}{\micro\meter}) black on the inside and white on the outside (see. Sec.~\ref{sec:muonveto}). The sheets are spanned in the interspaces between the steel profiles, providing a widespread but not complete optical separation of the two subvolumes. The allocation is depicted in Fig.~\ref{fig:tankandframe}. While large elements cover the lower and upper sections of the octagon sides, the middle section displays a chessboard pattern of small rectangular sheets and gaps for the PMTs. Several steel strings are spanned around the outer perimeter to provide additional fastening points. The PET sheets are folded-in several times at the edges and then fixed by stainless steel clamps. These are either attached to the strings or welded directly onto SF profiles. The horizontal PET sheets on the top of the frame will be mounted with a slight inclination to prevent trapping of air bubbles during the initial water filling of the WT.

\subsection{Acrylic Vessel}
\label{sec:av}

The container to hold and separate the LS from the surrounding water buffer has to be transparent, low in radioactivity, and sufficiently sturdy to permit continuous filling and exchange operations. Therefore, the LS vessel is made of acrylics that is a well proven material applied in many LS neutrino and some dark matter detectors and is used as well for the JUNO main detector.

\subsubsection{Mechanical design}

\begin{figure}[ht!]
    \centering
    \includegraphics[width=0.45\textwidth]{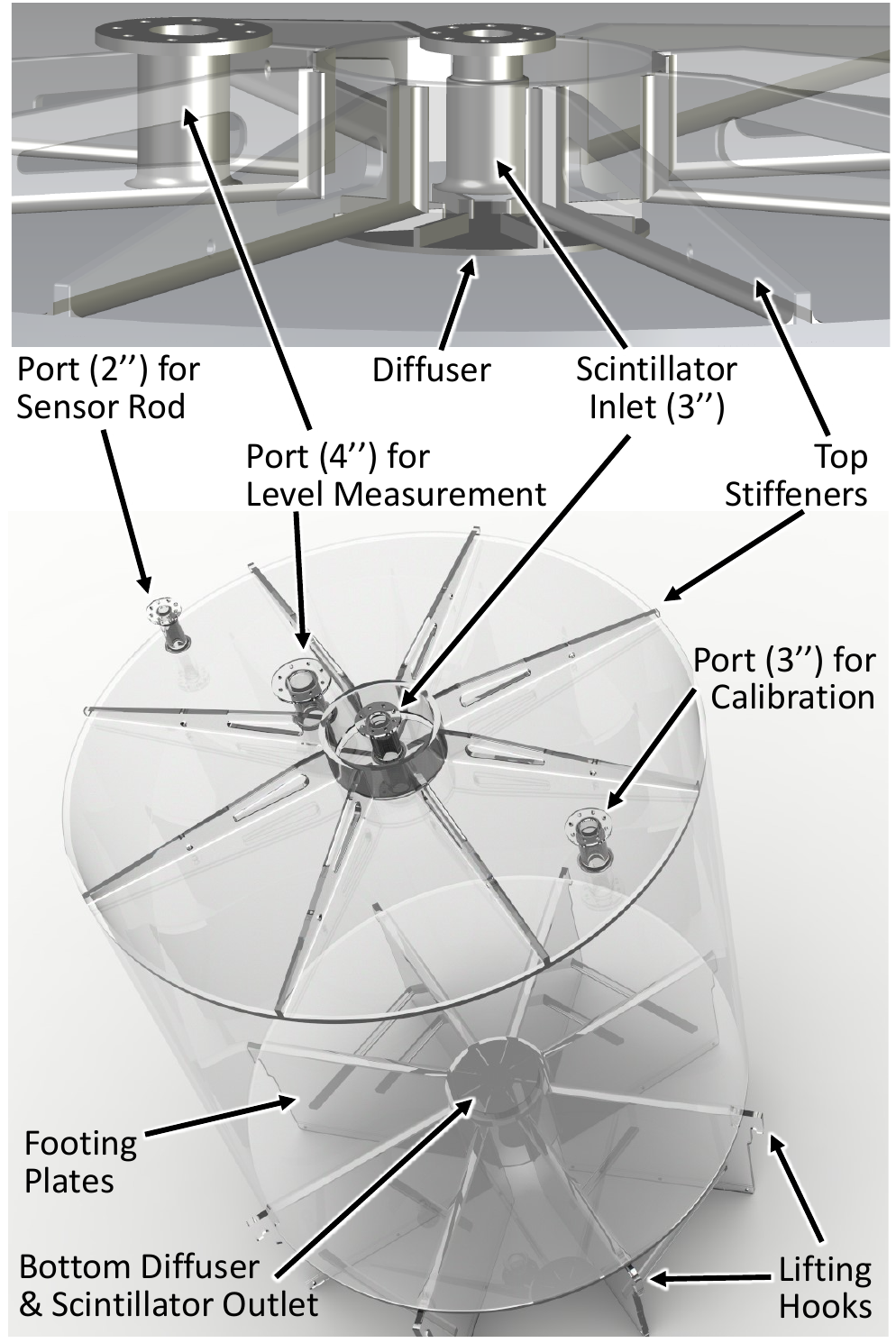}
    \caption{The Acrylic Vessel of OSIRIS: The lower panel shows a general overview, the upper panel a detailed picture of the LS inlet}
    \label{fig:AV}
\end{figure}

\paragraph{Bulk Dimensions} With an inner diameter of 3 meters and a height of 3 meters, the AV will hold a volume of 21\,m$^3$, corresponding to $\sim$18 tons of liquid scintillator. As shown in Fig.~\ref{fig:AV}, the cylinder mantle and faces are 30\,mm in thickness and re-inforced by straight and circular stiffening elements in order to increase the mechanical strength. While this is a relatively sturdy design compared to most neutrino detectors, it increases the mechanical robustness against inside-to-outside pressure differences of up to 30--50\,mbar, thus permitting comparatively easy filling and liquid exchange operations (Sec.~\ref{sec:av}).

\paragraph{Footing} The vessel rests on eight vertical acrylic plates bon-ded to the bottom of the cylindrical vessel. They are reinforced by two acrylic rings and several acrylic bars interconnecting the plates to increase stability. Each plate features a hook at the outer perimeter that are foreseen for lifting the AV onto the Steel Frame during installation. Along the lower rim of each plate, three 1'' diameter bore holes permit to fasten the AV to the Steel Frame by horizontal bolts. This holding mechanism is required to counter the buoyancy forces acting on the vessel when inner and outer volumes are filled with LS and water, respectively (almost 3\,t of lift are expected).

\paragraph{Ports} The AV provides several interfaces: In the center of both top and bottom lids, a connection to 3'' pipes for filling and draining the AV is foreseen (Sec.\ \ref{sec:lhs}). Next to the center, a 4'' pipe will be connected to permit measuring the liquid filling height in the AV by a laser system mounted on the top of the WT. A fourth pipe will connect to a 3'' socket at 1.2\,m radius to provide access for the ACU lowering calibration sources into the LS volume (Sec.\ \ref{sec:calibration}). Finally, a small opening (2’’ diameter) will be provided at the very periphery of the top lid for the installation of a rod equipped with temperature, pressure and liquid level sensors to monitor LS operations in the AV (Sec.\ \ref{sec:lhs}). All interfaces are realized by acrylic pipe sockets bonded to the AV lids and topped of with flat acrylic flanges. Their faces will be met by corresponding stainless steel flanges on the connecting pipes. The seal is provided by a custom-made double-O ring design.

\paragraph{Diffusers} In order to support temperature stratification of the LS during continuous filling mode (Sec.\ \ref{sec:lhs}), the central LS inlet and outlet are terminating within the AV volume not with an open end but are capped off by an acrylic disc of 500\,mm diameter acting as flow diffuser (fig.~\ref{fig:AV}). The discs are held in place by eight radial acrylic elements of 50\,mm in height. In this way, the stream of the LS is diverted from a vertical to a horizontal direction along the AV lids. The radial elements are meant to reduce as well circular motions (vortices) in the liquid. Dimensions are chosen to obtain low flow speeds and Reynolds numbers and thus mostly laminar flow of the LS. The functionality of the diffusers has been successfully tested in a scale 1:10 prototype.

\subsubsection{Production and Radiopurity}
Given the immediate contact of the vessel with the LS, its radiopurity is of uttermost importance for the sensitivity of the OSIRIS setup. Several quality assurance methods have been devised in order to ensure the compliance with the required purity levels of $^{238}$U, $^{232}$Th and $^{40}$K that has been specified to 1\ ppt level for both bulk and surface materials.

\paragraph{Production} 
The requested low-background acrylics for phy-sics studies will differ from the commercial products in the following protocols: the mold will be rinsed with ultrapure water before raw monomer injection; the polymerization will be catalyzed from a pure monomer without any additives for UV-resistance; thermal forming and bonding acrylics will be performed in a pure nitrogen environment; sanding and cleaning of the acrylic pieces will be performed inside a Clean Room of Class 10,000. The vendor (Goldaqua company) will record all surface contacts in a special diary. The quality assurance will be directly supported by physicists to reach the specified background level.

\paragraph{Bulk radiopurity} Acrylic samples are tested via ICP-MS (Inductively Coupled Plasma-Mass Spectrometry) that is a very practical and sensitive analysis method for discovering traces of radioactive isotopes. While the ppt-level abundances are too low to discover them directly in samples, a new method has been developed, where we adapt the dry ashing procedure in the acrylic sample and digesting it to solution in order to raise the concentration of radioactive element in the parallel samples. Together with ICP-MS, the corresponding detection limits of $^{238}$U and $^{232}$Th reach 0.32\ ppt and 0.70\ ppt, respectively.

\paragraph{Surface radiopurity screening} During the production and storage of the AV, additional surface contamination can be introduced by dust setting from the ambient air. Dust particles can be removed by rinsing the surfaces with deionized water. In the following, the wash water can be investigated for changes in electric resistance and light absorption properties, giving indirect evidence of the surface conditions. Studies are carried out to calibrate the contamination level of the water with the resistance and transmittance data, and can be related to the residual dust level on the surface by ICP-MS analysis. 

\subsection{Liquid Handling System}
\label{sec:lhs}

The Liquid Handling System (LHS) fulfills a range of requirements. It supports several operation modes for the filling and exchange of the liquids in the Acrylic Vessel (AV) and the Water Tank (WT) while maintaining appropriate liquid and gas pressure levels in all detector subvolumes to protect the detector from mechanical damage as well as the intrusion of radon. These tasks can only be achieved via a constant monitoring of the liquid and gas properties by a dedicated sensor system.

\subsubsection{Filling and operation modes}
\label{sec:modes}

While catering to a variety of needs for the operation of the detector, the functionality of the LHS can be roughly divided into three main operation modes:

\paragraph{Initial detector filling} Upon completion of the detector construction, the air inside the AV and WT volumes is replaced with ultrapure nitrogen to remove both oxygen (damaging the LS) and radon (introducing radioactivity). In a next step, both AV and WT are filled from below with high-purity water from a common water supply. The idea is to equalize liquid levels in the inner AV and outer WT volume, preventing the build-up of hydrostatic pressure differences. It also serves as a final cleaning of the inner surface of the AV from dust and radioactive contaminants. Finally, only the water inside the AV is replaced by LS that is now added from the top, floating on the residual water due to the lower density. The density difference leads as well to a potential difference in hydrostatic pressure that is held in check by the LS liquid level in the head tank.
\medskip\\

Unlike other LS detectors, OSIRIS sets special requirements to its LHS. Given that a large number of different scintillator batches is to be monitored during the commissioning and JUNO filling phases, the LHS is designed to permit exchange of the LS inside the AV without having to drain the outer water volume. There are two possible operation modes to achieve this: Batch and continuous-filling modes.

\paragraph{Batch Mode} In this mode, the new LS will be added through the top filling pipe but at temperature of at least +4\,$^\circ$C compared to the LS present inside the AV, while old LS is drained from the bottom outlet. The design rate of LS replacement is 1 ton per hour. Studies with a 1:10 laboratory prototype of the detector have shown that this temperature gradient will be sufficient to prevent larger intermixing between the upper layer of new and lower layer of old LS. The exchange of one full LS volume can thus be achieved within less than 1 day. In the follow-up, the new batch of LS can be screened for a period of several days to weeks, permitting to reach design-level sensitivity to U/Th contaminations (Sec.~\ref{sec:bipo_batch}).

\paragraph{Continuous filling mode} During the LS filling of the JUNO target volume, batch-mode operation does not provide a sufficient exchange rate of the LS to monitor all of the product LS for radiopurity. Moreover, stopping the filling to perform radiopurity monitoring of single batches creates extended dead periods in which a sudden increase in LS radioactivity cannot be discovered. Thus, an alternative replacement mode has been devised: without interruption, LS is continuously filled from the top at an increased temperature. The LS gradually cools down while traversing the full height of the AV within about 1-day's time and is then drained from the bottom. In order to maintain the temperature gradient inside the AV, a counter-flow of cold water is established on the outside in the water tank, enhancing the cooling speed of the lower layers of LS in the AV. With an exchange rate of 1~ton~per~hour, OSIRIS can monitor a constant fraction of $\sim$15\% of the product LS. As described in Sec.~\ref{sec:bipo_cont}, the U/Th sensitivity of this measurement mode crucially depends on the initial background level of radon introduced by the components of the JUNO and OSIRIS LS systems.

\begin{figure*}[t]
	\centering
	\includegraphics[width=1.00\textwidth]{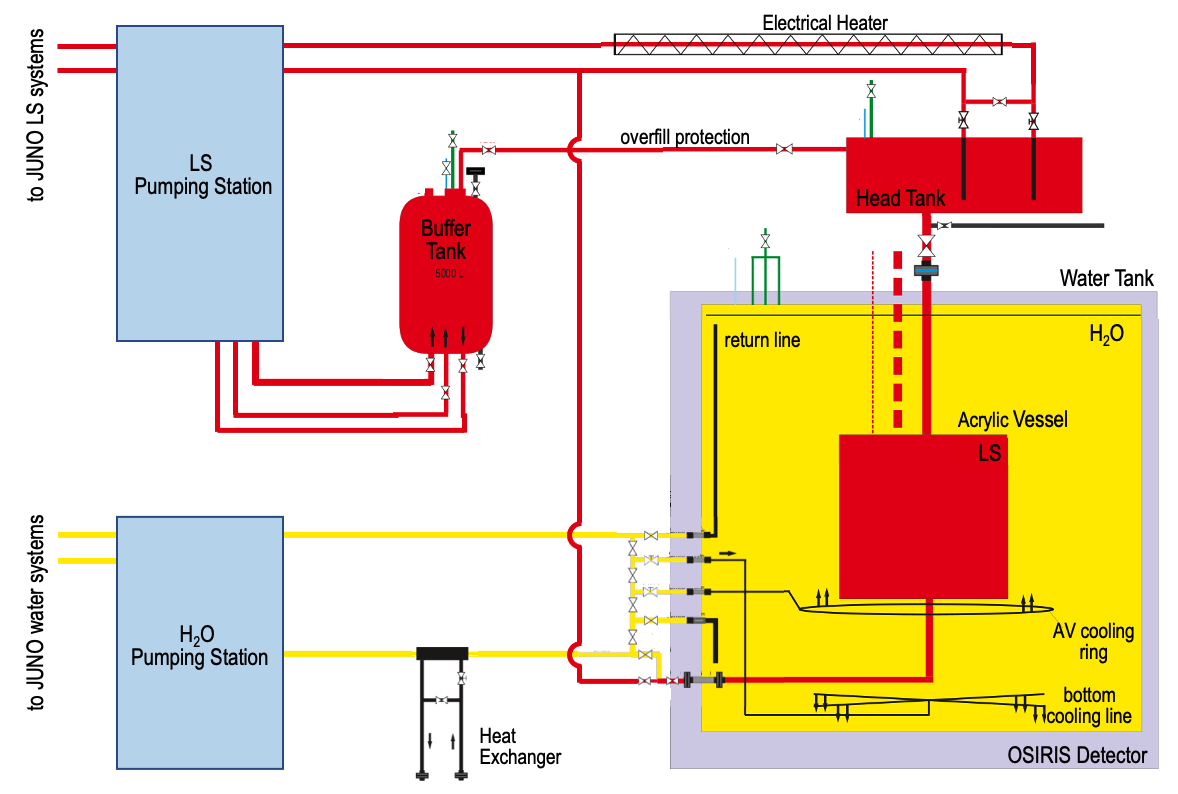}
	\caption{Conceptual layout of the OSIRIS LHS. The LS system (red) includes pumping station, buffer tank and head tank that permit to continuously exchange the LS within the Acrylic Vessel. An electric heating system permits to pre-warm the LS to establish a temperature gradient, causing LS stratification within the AV. The water system (yellow) supports this process by establishing a counter-flow of cold water along the sides of the AV, enhancing LS cooling}
	\label{fig:lhs}
\end{figure*}  

\subsubsection{LHS Layout}

Fig.~\ref{fig:lhs} shows a simplified diagram of the LS and water handling systems of OSIRIS. The system is realized as a bypass to the JUNO main LS line (not shown). LS and water systems can be operated in both a through-going and an internal loop mode, providing at least temporary independence from the external liquid supply. Here, we describe the main components.

\paragraph{Pumping Stations} Most of the components required for controlling the liquid flows of the OSIRIS LHS (magnetic centrifugal pumps, valves, flow meters and other sensors) are concentrated in the two pumping stations. LS passing from the JUNO main line into the OSIRIS LHS is redistributed by the LS pumping station (LS-PS), either to be temporarily stored in the Buffer Tank or to be forwarded to Head Tank and AV. In turn, LS extracted from the bottom of the AV is pumped out via a filtration stage in the LS-PS. The water pumping station is a simpler system that maintains the flow of water into and out of the WT with a single centrifugal pump. The nominal flow speed is 1~m$^3$/h for the LS system and 1--3~m$^3$/h for the water circuit. 

\paragraph{Buffer Tank} Directly connected to the LS input from the JUNO main filling line, a buffer tank with a storage capacity of 5\,t of LS provides independence from the operation status of the purification plants and CD filling systems. Even in case of an intermediary interruption of the external LS supply, the tank permits to maintain a constant inflow into the AV for several hours.

\paragraph{Head Tank} The stainless steel tank features a capacity of $\sim$100~l. It serves as a small LS reservoir above the AV that fulfills two important functions: during normal operation, the LS level will be kept at a defined height in the tank that is $\sim 50$\,cm above the water filling line of the WT to reduce the hydrostatic pressure difference between the inside LS and outside water volume surrounding the AV. Pressure equilibrium is achieved at the top of the AV to minimize the mechanical stresses. The tank plays as well a crucial role in the initial filling as it permits to carefully balance pressure levels when replacing water with LS inside the AV.

\paragraph{Electric Heater} In order to generate and maintain a temperature gradient within the LS inside the AV, the scintillator has to be heated prior to insertion at the AV top. This is achieved by a 10-kW electric heating tape attached to the pipe leading from the LS-PS to the Head Tank. The heating is temperature-controlled to less than 200\,$^\circ$C to avoid damage to the LS.

\paragraph{Water Cooling Loop} Water from the JUNO high-purity water system will be inserted into the tank by a loop pipe moun-ted just below the AV on the central Steel Frame to generate an upward flow of water along the sides of the AV. This water will be pre-cooled by a heat exchanger (10\,kW) in order to support the cooling of the LS. A second star-shaped system inserts water at the bottom of the tank to compensate any residual heat entering the tank through the floor. Warmed up water rises to the top of the WT and is extracted via a riser pipe. This open loop offers an effective way to remove radon added to the water by emanation from the detector components, while the incoming high-purity water features an expected radon level of 1\,mBq/m$^3$.

\paragraph{Nitrogen system} The nitrogen system of OSIRIS is fed by JUNO's high-purity N$_2$ supply. The system will receive N$_2$ gas at a pressure of 6--10\,bar. An initial pressure reducer provides N$_2$ at 0.5\,bar that will be used for the initial flushing of AV and WT volumes to remove oxygen from the system (Sec.~\ref{sec:modes}). A second pressure reducer supplies N$_2$ at +(5-10)\,mbar compared to surroundings. N$_2$ from this stage will be distributed to the WT, Head Tank, ACU and Buffer Tank to keep them under inert atmosphere. The slight over-pressure reduces the risk of ambient air entering the detector, while the use of high-purity N$_2$ to form blankets inside all tanks prevents diffusion of radon into the LS volumes.

\subsubsection{Liquid Level and Temperature Monitoring System}

\begin{figure*}[t]
	\centering
	\includegraphics[width=0.80\textwidth]{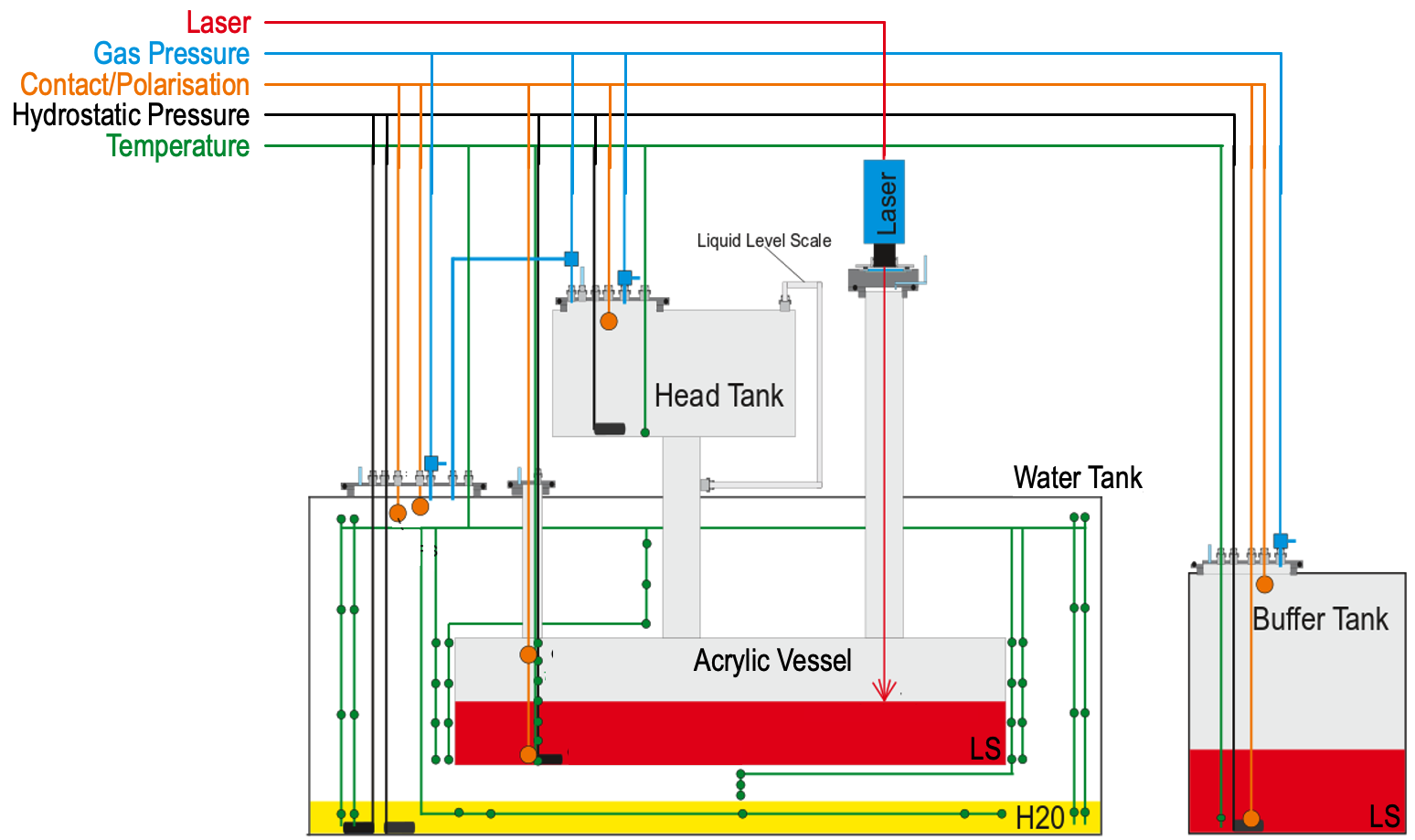}
	\caption{Schematic drawing of the LM-System with temperature sensor grid. The thermometers are indicated in green, the HPSs in black while the level switches are drawn in brown. The gas pressure meters are shown in blue. The red line shows the position of the IR laser level sensor}
	\label{fig:LMS}
\end{figure*}

The LHS sensor system serves two primary purposes. Operation of the complex system with several tanks and a direct connection to the external LS and water supplies requires a precise knowledge of the filling heights and flow rates to prevent damaging the system components. On the other hand, batch and continuous operation modes rely on a rather sophisticated control of the liquid temperatures inside the AV and WT, making close monitoring a necessity. An overview of the system is laid out in Fig.~\ref{fig:LMS}.

\paragraph{Level Monitoring System} The liquid levels inside all tanks are primarily monitored via hydrostatic pressure sensors inserted at the bottom of the respective liquid volumes (SISEN Automation, BST6600 PS3/6). They provide a relative resolution of $\pm$0.25~\% for the liquid levels. In the special case of AV, this measurement is complemented by an laser-based optical distance meter mounted on the roof of the WT (Laser Tech TruSense S310). The line of sight to the LS level inside the AV is provided by a dedicated 4'' stainless steel pipe connected to the AV top lid. The laser operates at an infrared wavelength of 905\,nm harmless to the PMTs and provides an accuracy of $\pm$10\,mm for the liquid level. For redundancy, all tanks are equipped with contact sensors at level heights critical to the filling process. Within the AV, this is realized with capacitive sensors that perform a measurement of the electrical polarizability of the contact medium and thus are able to differentiate between water and LS phases in the initial exchange operation (Rechner GmbH, Capacitive Sensors Series 80).

\paragraph{Temperature Monitoring Grid} Both {\it batch-} and {\it continuous-mode} operation benefit from a direct monitoring of the temperature profile inside the AV. Seven PT-1000 resistor thermometers ($\Delta T\sim0.1\ ^\circ$C) are arranged at regular distances along a stainless steel rod that is directly inserted into the LS volume close to the outer perimeter. This 'Thermorod' holds as well a hydostratic pressure and two capacitive sensors (see above). In the water volume, a grid of 16 thermometers arranged around the AV (supported from the Steel Frame by long radial stainless-steel rods) monitors the corresponding temperature profile of the upward water flow. This inner array is complemented by an outer thermometer array that provides information on potential external impact factors.    
\medskip\\
All sensors have undergone a HPGe screening to verify their compliance with the radiopurity requirements and the materials have been chosen for chemical compatibility with LS and water (mostly stainless steel and PTFE). Their cables are led from the WT via the roof Cable Flanges.

\subsection{iPMTs}
\label{sec:pmts}
 
Light read-out in OSIRIS is achieved by a total of 76 20-inch PMTs. The Inner PMT Array is formed by 64 PMTs arranged in four horizontal rings at different heights surrounding the AV and two rings of vertical PMTs below and above the AV lids (see Fig.~\ref{fig:osiris_layout}). Twelve further tubes of the same type form the Outer Detector.

The novel concept of the intelligent PMT (iPMT) combines the photomultiplier tube and the required electronics into a single device. The electronics -- high voltage supply, digitizer and control -- are mounted inside a stainless steel shell, which is located at the back of the PMT. The signal quality is improved by avoiding long cables between the signal source (PMT) and the digitization circuit. 

\begin{figure}[b]
    \centering
    \includegraphics[width=0.45\textwidth]{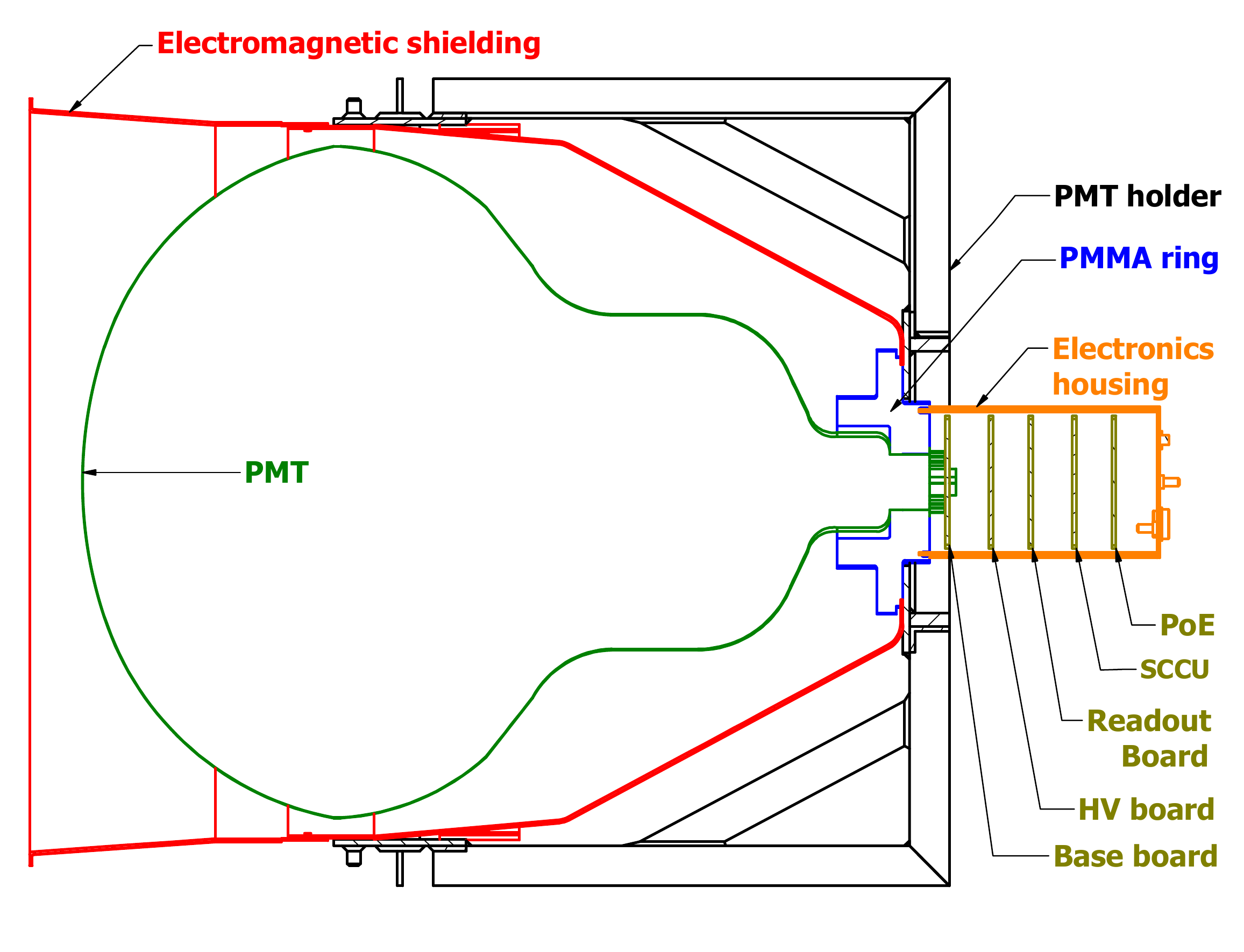}
    \caption{Cross section of one iPMT assembly}
    \label{fig:ipmt_crosssection}
\end{figure}
\subsubsection{iPMT Mechanics}

As shown in Fig.~\ref{fig:ipmt_crosssection}, each iPMT is composed of several parts that are assembled around the central PMT. The PMTs used for OSIRIS are \unit[20]{''} tubes of type R15343 by Hamamatsu. This type of PMT is based on a ``Venetian-blind'' dynode structure. The key parameters of the PMTs are summarized in table~\ref{tab:pmtproperties}.
\begin{table}[h!]
    \centering
    \caption{Properties of the OSIRIS PMTs}
    \label{tab:pmtproperties}  
    \begin{tabular}{l | r c l l}
        \toprule
         Parameter & Value \\ 
         \midrule
         Transit Time Spread & $2.63$ & $\pm$ & $0.16$ & \unit{ns} \\
         Dark Count Rate & $15.4$ & $\pm$ & $2.6$ & \unit[kHz] \\
         Peak to Valley & $3.21$ & $\pm$ & $0.20$\\
         \bottomrule
    \end{tabular}    
\end{table}

The electronics, consisting of a stack of five printed circuit boards (PCBs), is mounted directly to the back of the PMT. The basis for all mechanical mountings is a PMMA piece glued to the neck of the PMT. First of all, this PMMA piece provides an interface for glueing the stainless steel electronics housing. The electronics stack is immersed in oil which transfers the heat from the electronics via the stainless steel shell to the surrounding water. 

Each PMT is protected by a shielding cone from external electro-magnetic fields. The cones are based on aluminum (electric shielding) and amorphous metal (magnetic shielding) foils encapsulated by epoxy with carbon fibers for the inner PMTs respectively glass fibers for the OD PMTs. The front part of the shielding causes some optical shading for photons incident under large angles but is painted white to increase the chance that photons hitting the inner cone are reflected back on the photocathode.
The shielding is held in place by a stainless steel holder screwed to the PMMA piece. The holder supports as well the glass bulb of the PMT via a set of clamps arranged around the equator and provides fixing points for the interface with the Steel Frame. Horizontal PMTs are mounted at one of four holding plates arranged around the PMT equator, while vertical PMTs are fixated using a secondary steel ring at the PMMA piece.

\begin{figure}[t]
    \centering
    \includegraphics[width=0.45\textwidth]{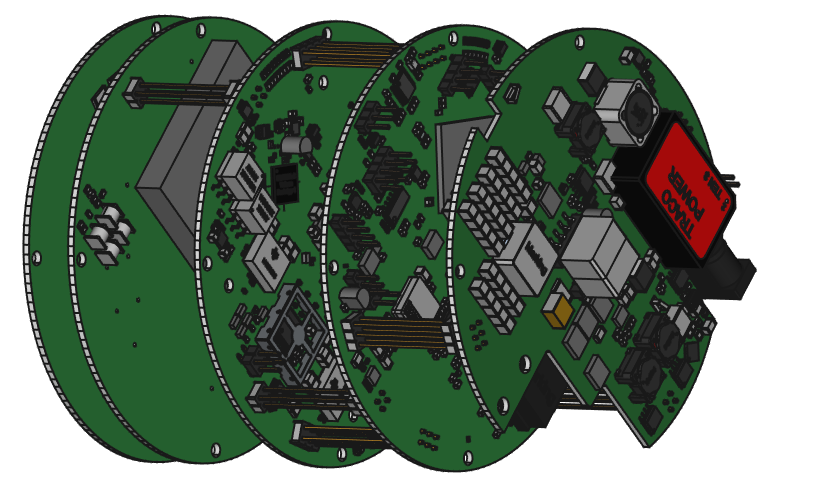}
    \caption{iPMT electronics stack. The PMT is mounted on the left side. From left to right, Base, HVboard, Readout Board, SCCU and POE. Spacing between the boards is not to scale}
    \label{fig:ipmstack}
\end{figure}
\subsubsection{Under Water Electronics}
As shown in Figure \ref{fig:ipmstack}, the electronics stack at the back of the PMT consists of five boards, each having a diameter of \unit[100]{mm}. Here we describe their function, starting from the lowest PCB closest to the PMT: 

\paragraph{Base} The voltage divider for the PMT is laid out according to the recommendations of Hamamatsu. The PMT is operated with a positive high voltage towards the anode, keeping the photo cathode connected to ground.

\paragraph{High Voltage Board} The board carries a custom-made high-voltage (HV) module, which converts \unit[24]{V} to a HV DC output suitable for operating the PMT \cite[p. 5]{Bellato_2021}. The board hosts as well the decoupling capacitors to strip the current signal from the HV supply.

\paragraph{Read-Out Board (ROB)} The ROB is the central board of the stack. On this board, the analog signal from the PMT is converted into a digital signal and processed further. The main component of the readout board is a System On Chip (SOC) unit (Xilinx ZYNQ 0720) which combines a FPGA with a dual core ARM processor. The FPGA part -- programmable logic (PL) -- is the hardware for handling the high speed signals of the digitizer. The other part, the processing system (PS), runs the software that controls the PL and forwards the data to the EventBuilder.

For synchronization purpose each ROB hosts a Clock-Data-Recovery (CDR) circuit, which is fed with a Manchester encoded data stream (synchronous link). The CDR recovers both clock and data from this data stream. The recovered clock is used as the reference clock for the digitizer. The data provides information about the synchronization of all iPMTs to a global timestamp.
\medskip\\
{\it VULCAN ASIC:} The conversion of analog to digital signal is done by the VULCAN ASIC. It has been developed by ZEA-2 at FZ J\"ulich to serve as the analog front-end electronics and the digitizer in a single chip. The ASIC provides three individual identical receivers, which can be highly configured using a JTAG interface \cite[p. 16]{PhDPavithra}. This interface also allows reading of multiple status registers.
Each receiver consists of an analog front-end and an ADC, which is configured to run at \unit[500]{MSps}. The three receivers are adjusted to cover three different signal amplitude ranges, the design values for the ranges are 0-10\,p.e., 0-100\,p.e., 0-1000\,p.e. \cite[p. 14]{PhDPavithra}.
This results in both a large dynamic range and high resolution at lower charges. 
All three receivers are sampled in parallel. The ASIC itself selects the one with the highest resolution and no saturation and sends the sample to the PL. This method provides the best amplitude resolution.
\medskip\\
{\it Signal Processing:} Inside the PL, the continuous digital data stream of VULCAN is serialized from two samples per clock cycle to four samples per clock cycle. The packets are continuously filled into a ring buffer. If the amplitude of a sample rises above threshold, a trigger is issued and the ring buffer is read out. Afterwards the data are converted into an Ethernet-packet and transferred via a buffer and the Ethernet link to the EventBuilder (\ref{sec:daq}).

\paragraph{Slow Control and Configuration Unit (SCCU)} The SCCU hosts a STM32 microcontroller and an four-port Ethernet switch (only three ports are used). This switch provides a port as an uplink to the Surface Board. A second port is connected to the ROB while a third port is interfaced by the microcontroller. The latter runs a firmware that handles the Xilinx XVC protocol for the JTAG access and custom protocols for accessing two I2C ports and three UARTs. One of the UARTs is connected to the HV module via RS485 and two are connected to the ROB.

\paragraph{Power Over Ethernet (POE)} This board hosts a POE class 4 powered device controller, which is followed by two galvanic isolated DC/DC converters. While one creates a dedicated \unit[24]{V} supply for the HV module, the other supplies a \unit[5]{V} net for the remaining electronics.
\medskip\\
When all boards are in normal operation, the stack consumes about \unit[11]{W} at the POE side. Since the stack is surrounded by oil all contacts on the boards are soldered. Plugs would carry the risk of oil creeping in and thus decreasing the quality of the electric contact. During assembly, the boards are first connected by bottom entry connectors for testing. Soldering is done as a last step without unplugging. 

\subsubsection{Surface Electronics}

The connection between the iPMTs and the Surface Electronics is established by a separate Ethernet cable for each iPMT. The length of the cable depends on where the iPMT is placed inside the Water Tank (17m for the iPMTs at the top of the detector, 22m for the center rings and 25m for the iPMTs at the bottom). The individual cables exit OSIRIS via the Cable Flanges on the top lid of the Water Tank. 

The cable consists of four wire-pairs, which are split up in two specially developed Surface Boards (each servicing up to 48 iPMTs). The two Ethernet wire-pairs are directly forwarded to a POE switch, the two other pairs are used for the synchronization.

It is mandatory for the event reconstruction in OSIRIS that all iPMTs are synchronized to a single global time stamp. This time stamp is created by one Surface Board in regular intervals and broadcast to the other Surface Board as well as to the connected iPMTs via the synchronous link.

The iPMT is running on the global reference clock via the CDR. Once the iPMT is powered, the local time stamp of the iPMT is set to the global time stamp. In order to monitor the synchronization, the local and global timestamps are compared. 
The designed accuracy of the timestamps between two arbitrary iPMTs is better than \unit[1.5]{ns}.

For creating a trigger signal for the laser and LED calibration system, each Surface Board is equipped with two outputs. These outputs create a pulse and at the same time a timestamp is sent to the EventBuilder.

\subsubsection{Slow Control}
The slow control system of OSIRIS is based on the Experimental Physics and  Industrial Control System (EPICS) \cite{EPICS} version 3{.}14. By this, the slow control system is compatible with the JUNO Detector Control System (DCS) \cite[p. 240]{JUNOCDR}. For an easy accessibility of the OSIRIS slow control, there will be a graphical user interface based on {\it caqtdm}.
\medskip\\
{\it iPMT Slow Control:}
The iPMTs provide many slow control data sources. Each iPMT provides about 100 slow monitoring variables. 
These variables include data such as measurements from I2C sensors (voltage, current, temperature etc.), the status of the PS or status registers from VULCAN. These status registers can not be logged in real time, but polling them several times per minute might be useful for further analysis, for example waveform reconstruction.
\medskip\\
{\it Ethernet switches:} The POE-Ethernet switches are the backbone of the OSIRIS detector. They are providing power to each iPMT. For OSIRIS managed switches are used. They provide a SNMP interface, which allows remote control of the Ethernet as well as power. This functionality is mapped to be available in EPICS. 
\medskip\\
{\it Integration of calibration systems:}
The slow control software of both calibration systems, ACU and fiber system (see chapter \ref{sec:calibration}) is based on LabVIEW 2015. Both systems can be used via independent GUIs. 
They are included in EPICS via the LabVIEW shared variables engine, which provides remote control.


\subsection{Calibration Systems}
\label{sec:calibration}

The calibration of the OSIRIS detector is performed based on two subsystems. A {\it fiber system} coupled to a very fast laser is used for timing and charge calibration of all PMTs (Sec.~\ref{sec:fibersystem}). The light pulses are distributed from the laser via optical fibers to insertion points placed at several locations on the Steel Frame. This allows a precise relative timing and single photo electron (s.p.e.)~charge calibration of all PMTs. Additionally, a {\it source insertion system}, placed at the top of the detector, is used to lower a selection of weak radioactive gamma sources as well as an LED directly into the LS volume (Sec.~\ref{sec:sourcesystem}). While the LED provides redundancy for the timing and charge calibration, the gamma sources are used to calibrate energy and position reconstruction.

\subsubsection{Fiber System}
\label{sec:fibersystem}

The two main tasks of the Fiber Calibration System are a precise calibration of the relative timing of iPMTs on a sub-nanosecond level and the equalization of s.p.e.~charges. Several design requirements have been considered in the development of the system:
\begin{itemize}
    \item A pico-second laser has been chosen as light source to achieve the required time resolution. Both time and char-ge calibration require illumination of the PMTs in the sub-Poissonian regime. Thus, the aimed-for light intensity is $\mu=0.01$ p.e. per pulse. In addition, differences in light running time have to be avoided using fibers of well-defined length, with same-length fibers for groups of PMTs. Remaining inequalities should be negligible compared to the TTS of the PMTs (Tab.\ \ref{tab:pmtproperties}). 
    
    \item The duration of a daily calibration run should not exceed 10\,min: starting from a TTS of 1.1\,ns ($1\sigma$, cf.~Tab.~\ref{tab:pmtproperties}), about $2\times10^3$ s.p.e.~pulses are required to reach a relative timing alignment of 25\,ps ($1\sigma$) between channels. Due to the low average light intensity, about $2\times10^5$ pulses per PMT are required. With a maximum trigger rate of 10\,kHz, this leads to a measurement time of 20\,s
    per diffuser and a total of 8\,min for a complete run. The corresponding accuracy expected for the PMT charge calibration is at the $7\times10^{-3}$ p.e.~level.
    
    \item All internal parts of the system need to comply with the radiopurity limits as well as guarantee a reliable long-term operation. A flexible hardware design (e.g.~linearity checks of the PMTs) is needed, too, to provide the opportunity to adjust to further requirements in a future physics program of OSIRIS. 
    
    \item The system must provide an interface for remote control, i.e.~it has to be integrated into the slow control/EPICS system of OSIRIS.
\end{itemize}
Based on these requirements, the system features the following components:

\paragraph{Light source} The pico-second Laser is a PiL042X from A.L.S. GmbH. It generates pulses 80\,ps in length with a wavelength of 420\ nm and a maximum repetition rate of 20\ MHz. The light intensity can be tuned and a standard FC/PC fiber coupling is provided.

\paragraph{Light injection points} The positions and number of light injection points have been optimized based on the the detector simulation. They guarantee a uniform illumination of all PMTs with light intensities varying no more than a factor of two between individual PMTs. To achieve this, a total of 24 injection points with diffusers will be installed on the Steel Frame: eight inward-facing diffusers distributed at regular intervals at the height of the AV equator, two times four vertical diffusers below and above the AV to illuminate the PMTs on the opposite side of the AV, and another eight distributed in the Outer Detector (see chapter \ref{sec:steelframe}) 

\begin{figure}[h!]
    \centering
        \includegraphics[width=0.48\textwidth]{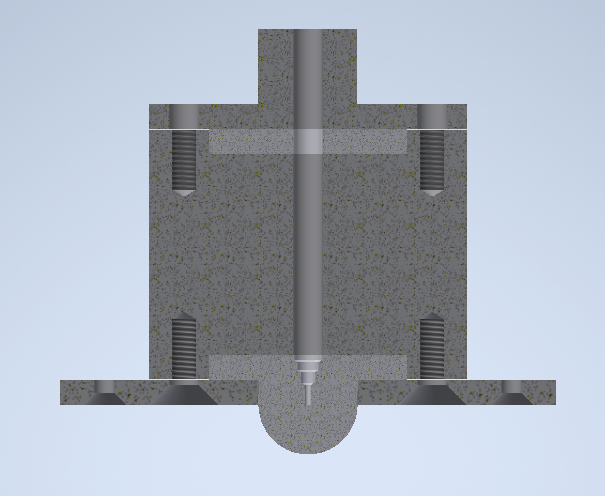}
        \caption{Cross section of the diffusor capsules. Sections displayed in light grey are made of PTFE, dark grey is 316L stainless steel}
    \label{fig:fiber_sys}
\end{figure}

\paragraph{Diffuser capsules} At the light injection points, PTFE diffuser bulbs have been attached to the ends of the individual fibers to create a homogeneous light field. The ferule tip of the fiber is placed in the center of the PTFE hemisphere, which results in a homogeneous radiation field in the range of 70$^\circ$ from the central axis of the hemisphere. The bulbs are encapsulated in a stainless steel shell. The necessary sealing is realized in two ways: Firstly, the fiber is glued into the back of the capsule using an epoxy (Masterbond EP30-4). Secondly, an additional PTFE disc seals the top plate of the diffuser capsule. A schematic drawing is presented in figure~\ref{fig:fiber_sys}.

\begin{figure}[h!]
    \centering
        \includegraphics[width=0.48\textwidth]{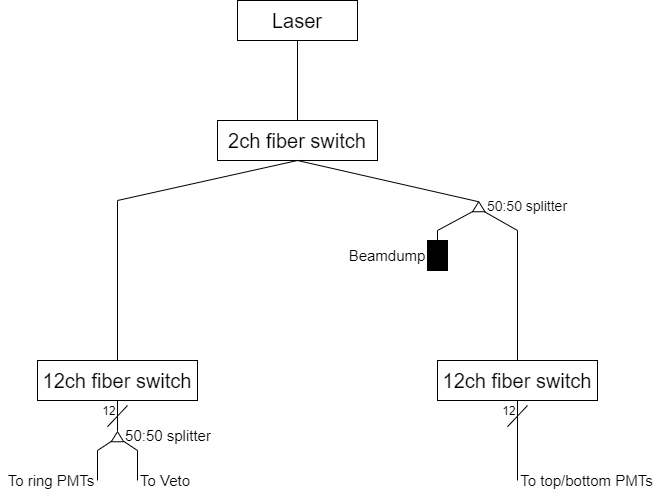}
    \caption{Routing scheme of the light distribution system for the calibration Laser}
    \label{fig:fiber_routing}
\end{figure}

\paragraph{Light distribution system } A cascade of optical switches and 50:50 fiber splitters will distribute the light from the single laser source to the light injection points. As illustrated in fig.~\ref{fig:fiber_routing}, the channels of the system are divided into two groups, covering the top/bottom and ring/veto diffusors. Sin-ce the Outer Detector and Inner Detector are optically separated, such a split enables to illuminate veto and ring diffusors simultaneously, leading to a reduction in calibration time.
\medskip\\
Laser and distribution system will be housed in a 19" rack mount on the WT roof. The active elements can be controlled via the slow control system using a standard RS232 interface. Additionally, a LabVIEW control software is avai-lable, providing all system parameters as EPICS IOCs for remote operation.

\subsubsection{Source Insertion System}
\label{sec:sourcesystem}

The Source Insertion System is mounted to the top of the Calibration Flange on the roof of the OSIRIS Water Tank. It is based on an Automated Calibration Unit (ACU), which has been provided by the Daya Bay collaboration \cite{LIU201419}, and then refurbished to suit OSIRIS’s needs. A 3'' stainless steel pipe provides a connection to the AV and permits to lower sources directly into the LS volume. Here we describe the main components.

\begin{figure}[t]
    \centering
    \includegraphics[width=.45\textwidth]{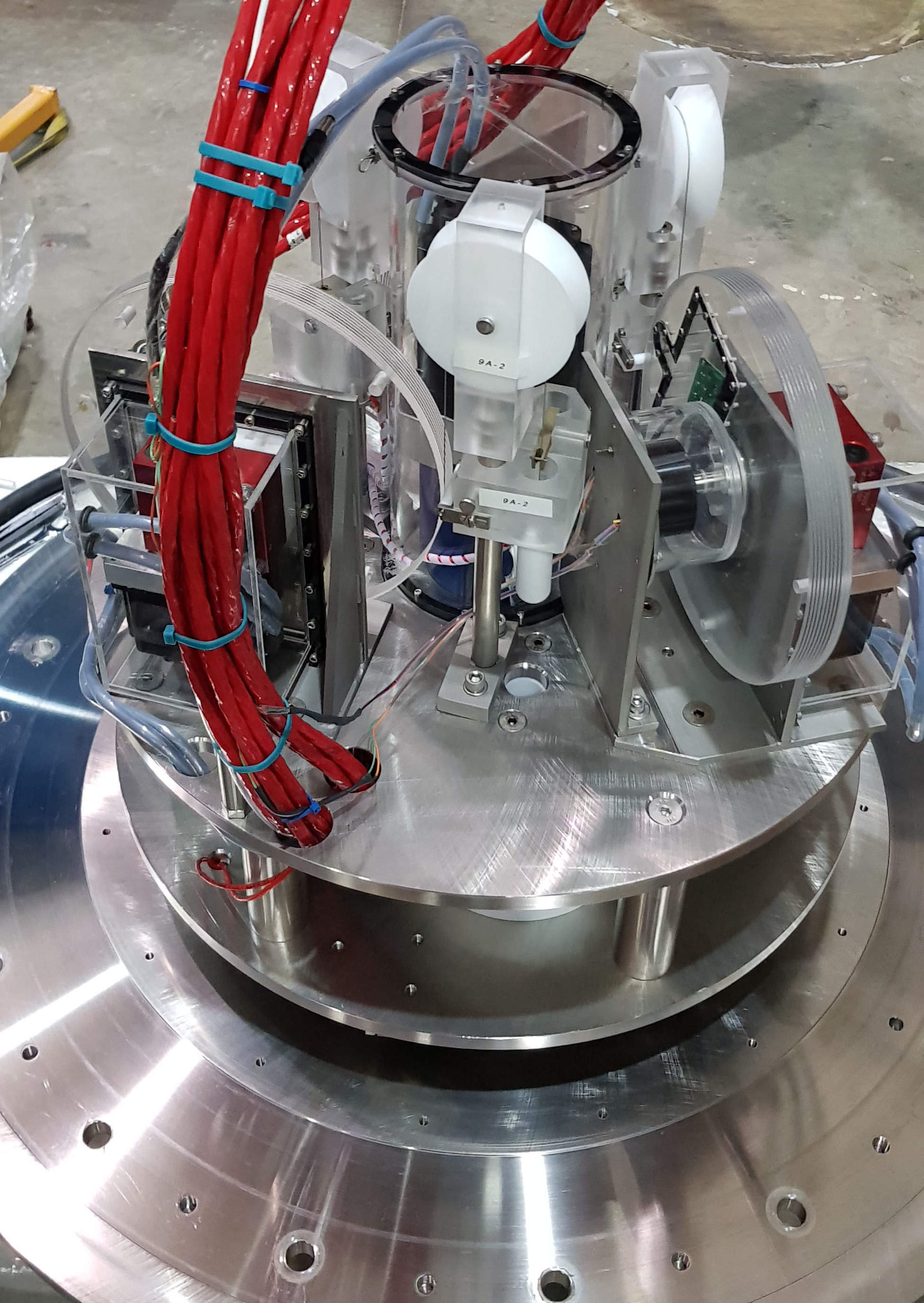}
    \caption{The Automated Calibration System (ACU) with the bell jar removed. Three acrylic winches are mounted to a turntable, allowing to lower sources down into the detector through a 1-inch hole in the bottom plate }
    \label{fig:source:acu}
\end{figure}

\paragraph{ACU mechanics} The ACU interior is shown in Fig.~\ref{fig:source:acu}. A turntable supports a revolver-like structure of three acrylic winches that each hold a several meter-long string with a source attached at the end. One of the winches carries an LED used for timing and charge calibration of the PMTs. The two others carry radioactive gamma sources to calibrate energy and position reconstruction and to monitor the light yield of the LS. All ACU components have been selected for low radioactivity and are chemically compatible with the LS (PMMA, PTFE and stainless steel). During operation, the ACU is covered by a bell jar that will be flushed with ultrapure nitrogen to prevent contamination of the LS with oxygen or radon.

The bottom plate of the ACU provides a 1'' port with an attached gate valve that is connected to the pipe leading down to the AV. The sources are lowered by stepper motors that unroll the strings on the winches. The motors are equipped with gear boxes and a combination of load cells and limit switches to ensure an accurate and safe positioning. The entry port is located at a radius of 1.2\,m from the central detector axis. This insertion axis has been chosen to maximize the variation of the detector response at different heights, as sampled by various sources. The detector simulation predicts a 10\,\% variation of light collection and thus energy response over the fiducial volume. Acquiring a selection of varied calibration measurements along the axis allows for maximal handles to tune the simulation, that will in turn be used to extrapolate the spatial dependence of the energy response to the entire LS volume.

\paragraph{LED source} The source contains a blue LED (\SI{435}{\nano\meter}) able to emit fast low-intensity pulses for time and charge calibration of the PMTs. To emit isotropic light, the LED is inserted into a PTFE diffuser ball. Both are enclosed in an acrylic capsule. The LED is connected to a fast pulser circuit based on a design by Lubsandorzhiev $\&$ Vyatchin \cite{Lubsandorzhiev}. It allows to pulse the LED with up to \SI{3}{\kilo\hertz} and an adjustable signal intensity, creating pulses of a minimum duration of $\sim$7\,ns (FWHM).

When driven at low intensities, the LED will provide a cross-check to the s.p.e.-based results of the fiber system (Sec.~\ref{sec:fibersystem}). In addition, the tunable intensity offers the chance to investigate the PMT response to higher p.e.~occupancies. This is important since the average signal per PMT expected for OSIRIS is $\geq$2 photo electrons. The LED light intensity must be chosen with some care since the off-center position of the sources in the LS volume means that nearby PMTs may receive a factor of 10 higher intensity than the farthest tubes. In the s.p.e.~mode, the intensity will be chosen to achieve an occupancy range from 5\% to the closest to 0.5\% the most remote PMTs. According to preliminary Monte-Carlo studies, 50 to 100 thousand LED pulses will be sufficient to obtain an average relative timing uncertainty of $\leq$1\,ns  and a charge calibration uncertainty of (3-5)\,\% for the furthest PMT. Based on the high repetition rate, such a measurement can be completed within a few minutes.

\begin{figure}[t]
    \centering
    \includegraphics[width=0.48\textwidth]{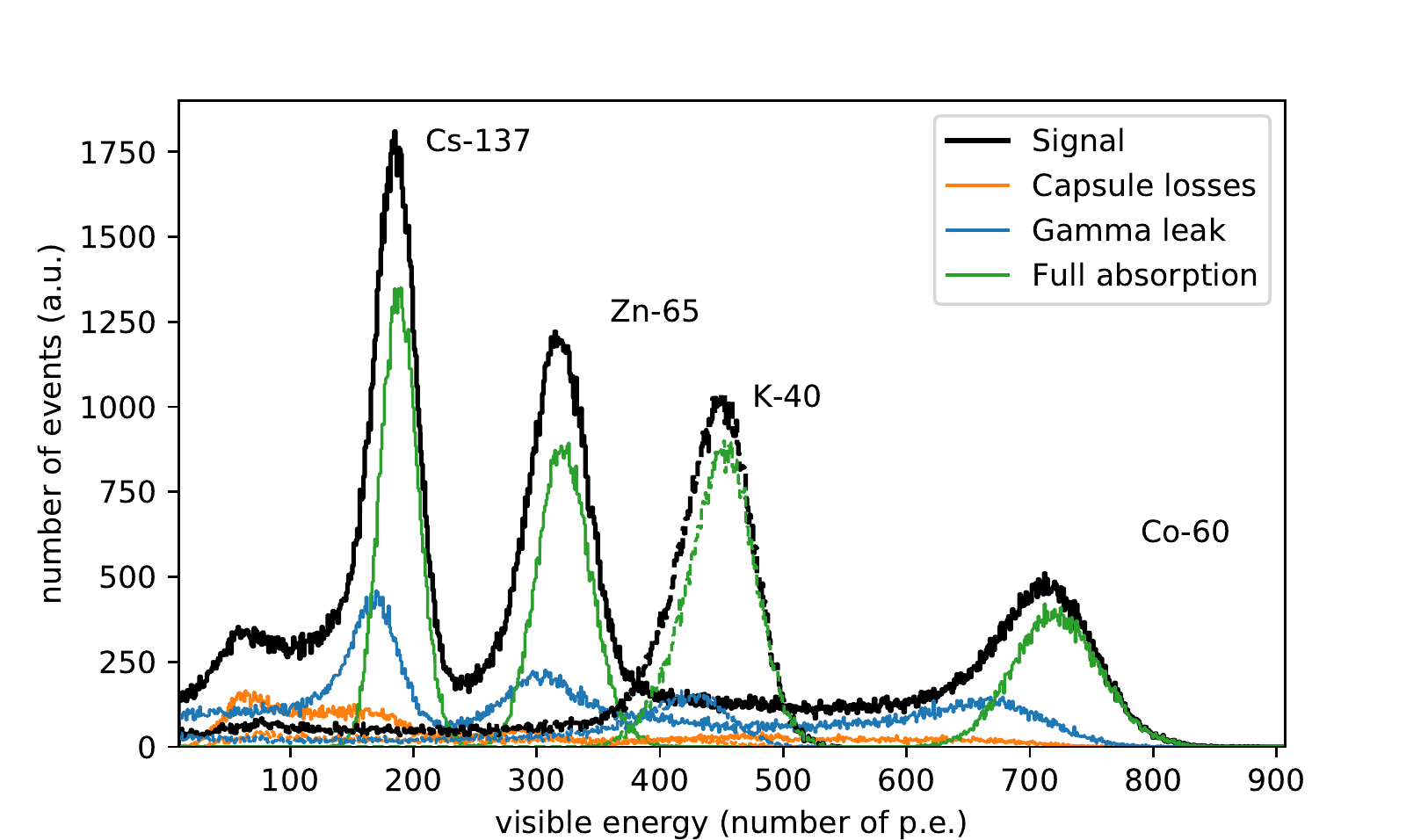}
    \caption{Simulation of the visible energy spectra (in number of photo electrons) of the gamma sources placed off-axis at a radius of 1.2\, and 1.2\,m over the detector center. The dashed line represents the signal of the $^{40}$K capsule, the solid line of the multi-gamma source. The absorption peaks of the individual isotopes are clearly discernible. The observed spectrum can be understood as a combination of events with full absorption in the LS ({\it green}), with energy loss in the capsule ({\it orange}) and with energy loss from spill-out from the AV volume ({\it blue}).}
    \label{fig:source_spectrum}
\end{figure}

\paragraph{Multi-gamma source} This is the primary source for the calibration of event energy and position reconstruction. The PTFE capsule contains a combination of three radioactive isotopes: $^{137}$Cs, $^{65}$Zn, and $^{60}$Co. With a combined activity of several \SI{}{\kilo\becquerel}, this source is used to probe the energy range from \SIrange{0.66}{2.5}{\mega\electronvolt}. The corresponding visible energy spectrum is shown in Fig.~\ref{fig:source_spectrum}. The $^{137}$Cs produces a \SI{0.66}{\mega\electronvolt} gamma, the $^{65}$Zn a \SI{1.12}{\mega\electronvolt}, and the $^{60}$Co produces two coincident gammas with energies of \SI{1.17}{\mega\electronvolt} and \SI{1.33}{\mega\electronvolt}. The lines cover the energy range crucial for the detection of Bi-Po signals (Sec.~\ref{sec:bipo_cuts}).

\paragraph{Potassium source} The capsule contains natural potassium resulting in $\leq$\SI{1}{\becquerel} of $^{40}$K gamma emission ($^{40}$K abundance: 0.0117\%). The activity is sufficiently low to keep it inside the LS volume during normal operations. This allows for a continuous monitoring of the LS light yield via the $^{40}$K peak position. Shifts may indicate a change in LS composition and would be reported to the JUNO filling team.

\paragraph{CCD system} An infrared camera system will be installed on the Steel Frame to support the source calibration system. Two inward-facing cameras equipped with LED lights will monitor and cross-check the position of the capsules inside the detector, checking especially for horizontal deviations from the calibration axis. The design accuracy is $\sim$\SI{3}{\centi\meter}.

\subsection{Muon Veto}
\label{sec:muonveto}


In order to identify muons and veto signals from associated secondary particles in the liquid scintillator, the water shield volume is utilized as an active Cherenkov veto. 
It is instrumented with 12 20''-PMTs to detect Cherenkov photons from muons in the water volume. 
PMTs, electronics and mechanical mounting are identical to those of the inner array. The magnetic shielding is realized by less radiopure glass fibers instead of carbon fibers. 

\paragraph{PMT arrangement} As displayed in Fig.~\ref{fig:simu_render}, eight PMTs will be placed on the bottom of the steel tank facing upward and four PMTs will be installed horizontally on top of the Steel Frame to cover the entire water volume. The positions have been optimized using the OSIRIS simulation framework (Sec.~\ref{sec:simulation}), maximizing the detection efficiency for muons that pass only the water volume. All muons crossing as well the LS volume inevitably feature a sufficient track length for efficient detection ($> \unit[99]{\%}$) and can be clearly identified by their high scintillation signal.

\paragraph{Reflectors} To increase the number of detected Cherenkov photons in the veto system, the side walls and the floor of the veto volume will be covered with white Tyvek foil which is attached to the steel tank liner. 
As the increased reflectivity would have a negative impact on the time resolution of the ID, the inner and outer PMT systems are optically separated by bicolored PET sheets fixed to the PMT holding frame. While white on the outside, the inner sides of the sheets are black to avoid reflections interfering with the time response of the ID.
This separation is not completely hermetic but simulations show that a coverage as indicated in Fig.~\ref{fig:tankandframe} is sufficient to prevent cross-talk on a level that would diminish the detector performance. 

\paragraph{Muon trigger} Based on simulations, three different trigger schemes have been elaborated\footnote{The finally utilized trigger condition, depends on the DAQ properties}. For each, a set of trigger conditions has been optimized to minimize false triggers induced by coincident PMT dark counts while maximizing the muon detection efficiency. The trigger schemes differ in their requirements for multiplicity of hit PMTs and the minimum number of detected photons within the trigger window. In the most sophisticated scheme, also the relative positions of the hit PMTs are considered to exploit the locally restricted photon emission along the muon track.

\paragraph{Expected performance} In order to determine the required muon veto performance, an extensive set of cosmic muon events and their secondaries was simulated. Neutrons and radioisotopes from muon spallation in the LS can pose a correlated background to Bi-Po coincidence searches. Applying the corresponding selection cuts (Sec.~\ref{sec:bipo_cuts}), the expected rate of \ce{^{214}Bi}-Po-like background events is 0.1 per day in the LS volume.

In the presented veto setup, the muon detection effici-ency is expected to be 90-97\%, depending on the chosen trigger model. The resulting background rate is equivalent to a uranium contamination of the LS on the level of $10^{-18}$\,g/g, permitting the screening of LS samples to similar contamination levels. The performance of the muon veto has been shown to be robust under PMT failure. The veto system can handle a loss of up to 4 PMTs (depending on the exact location of the tubes) without a significant loss in detection efficiency.

\subsection{Online DAQ}
\label{sec:daq}
\begin{figure*}
    \centering
    \includegraphics[width=0.8\textwidth]{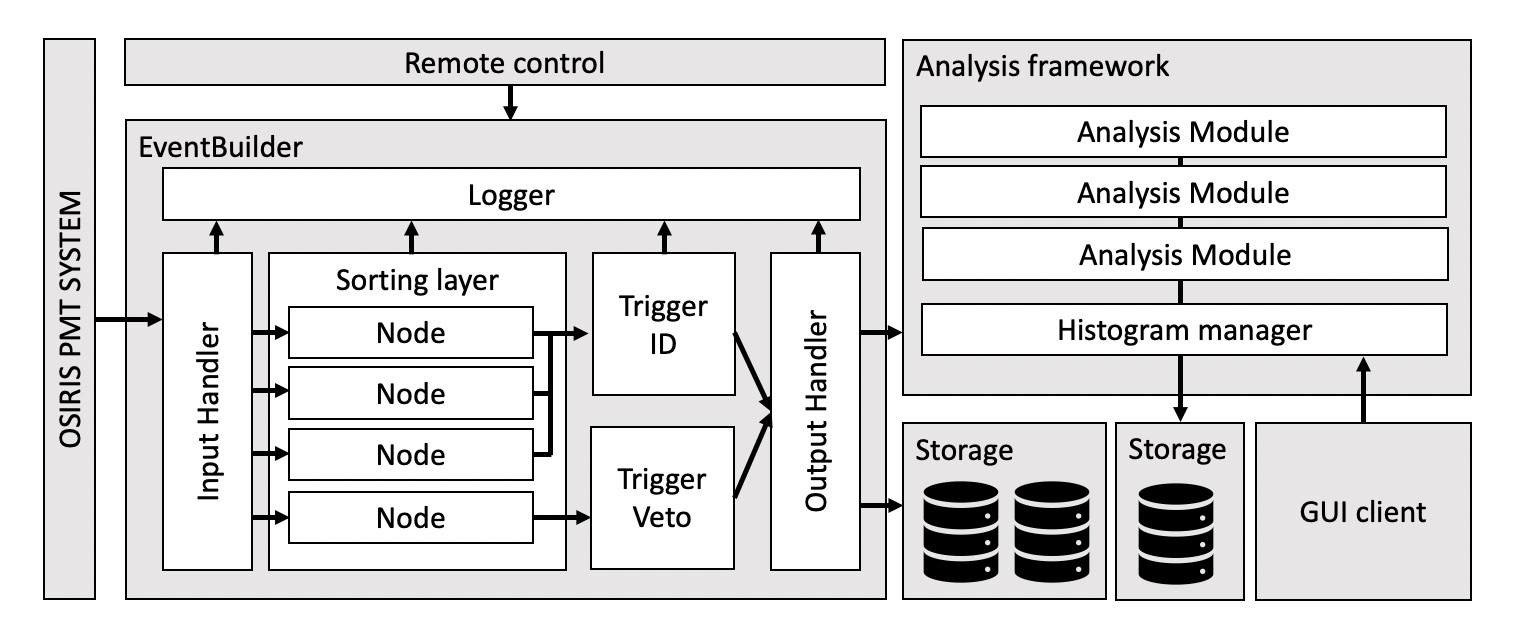}
    \caption{Schematic overview of the components and data processing in OSIRIS DAQ software}
    \label{fig:daq_scheme}
\end{figure*}

The structure of the OSIRIS DAQ is adjusted to the photo-detection system based on the novel iPMTs. Since the sensors are self-triggering, every time a single iPMT meets the internal trigger condition a data packet will be sent out over the network. That allows, in principle, running OSIRIS in a trigger-less mode. On the other hand, the data flow is heavily dominated by the dark counts in which the time-correlated PMT hits of physics events have to be identified and extracted. This task is performed by the online trigger that is part of the functionality of a software called {\it EventBuilder}. Furthermore, it is central to the functionality of OSIRIS that changes in the event rates and especially Bi-Po coincidence signals are monitored in real time. Therefore, event reconstruction and basic analyses have to be performed online. To do that, the triggered events are sent over the network to a subsequent analysis framework that takes care of online analysis and monitoring. The schematic overview of the components are shown in figure \ref{fig:daq_scheme}.

\paragraph{EventBuilder layout}
In general, the trigger decision in the EventBuilder relies on three crucial stages: 1) an interface between the iPMT-system and EventBuilder takes care to read in all data packets sent over the network, 2) the data packets are time-sorted according to their time stamps (including preliminary corrections related to cable lengths and calibrations) and 3) coincidences in the data flow are identified based on the data-packet time stamps. When the trigger condition is met all the corresponding data packets are bunched together to form a raw event. The hit data is written to disk and forwarded to the online analysis framework. All the relevant tasks (reading in, time-sorting, trigger-finding, etc.) are implemented as parallel threads in the EventBuilder software. For this, the data packets are intermediately stored and transferred forward via data containers accessible by the relevant threads. The containers are implemented as single-producer-single-consumer-lock-free queues to guarantee a maximum reading and writing speed. The most CPU-inten-sive task is the time-sorting of the incoming data packets. To reach sufficient performance, several sorting nodes have been arranged into a sorting layer that distributes the incoming data packets for parallel sorting and later on merges the separate threads into a single (subsystem dependent) data flow for the trigger search. Sorting itself is implemented using a binary search tree-approach. 
 
 \paragraph{Trigger building}
The trigger building thread accesses the time-sorted data flow and performs a search of coincident PMT hits meeting the trigger conditions. Different trigger modes have been implemented, including a standard physics mode, a calibration mode and several modes to debug and inspect the performance of each software layers of EventBuilder. The standard physics trigger is based on the coincidence of five PMT time stamps within a time  window of 70\,ns. All data packets within a pre-trigger window of 200\ ns before  and 800\ ns after the trigger are selected to form the raw event. After an event is constructed, it is written to disk and forwarded to the online analysis framework. It has to emphasized that trigger-building in software allows a flexible tuning of the trigger conditions and logic implementations at later stages of the OSIRIS programme. 

\paragraph{Performance evaluation}
The performance of the EventBuil-der software and the considered hardware has been evaluated by simulating the data acquisition chain. Two PCs were connected together via 10 Gbit/s network cards. The generator PC representing the iPMT system produced data packets and sent out a data flow to the second EventBuilder PC running the DAQ software. The throughput of the DAQ chain was tested by running the setup with 76 fake data generators with dark count rates taken from the PMT data sheets of the manufacturer. The test results demonstrate that the EventBuilder can process data rates that are more than a factor of 2 higher than those expected from PMT dark counts. The hardware required is an 8-core CPU (Intel i9 9900KF) with 128 GB memory and a 1-TB SSD storage configured to run in RAID-1 (mirroring) to ensure data safety.

\paragraph{Online/offline analysis and GUI}
The online analysis and monitoring framework of OSIRIS is based on the RootSorter online/offline analysis toolkit developed for accelerator experiments (ANKE and others) at COSY \cite{RootSorter}. The core part of the RootSorter package provides a comprehensive list of functionalities typically needed, such as the communication method (TCP/IP), templates for data handlers, interfaces to different kind of parameter objects (geometries, calibration tables, run tables etc.) and an online monitoring client. It provides also a modular environment to run different kinds of analyses on the incoming data. The RootSorter package is linked against ROOT to access a vast collection of powerful data analysis methods and tools. A basic user application for OSIRIS has been constructed by implementing the required data structures and data handlers to receive the data sent from the EventBuilder. Based on this, a GUI for simple online monitoring has been set up. The analysis and monitoring framework is under constant development and full online event reconstruction, search for coincidence signals (e.g.~Bi-Po) etc.~are still to be implemented for the OSIRIS commissioning.

\section{Detector Simulation}
\label{sec:simulation}

The OSIRIS simulation framework has been devised to study the detector response to the different signal and background sources and thus optimize the design of the experimental setup. It utilizes the particle simulation package GEANT4 \cite{Agostinelli2003} (Version 10.2.3) and is linked against ROOT\cite{Antcheva:2009zz} and CLHEP\cite{Lnnblad1994} to allow efficient data structuring and output formatting as well as conversions and calculations in the physics context. This section summarizes the design of the simulation framework, discusses the expected performance of the OSIRIS detector and introduces the main background contributions for the sensitivity studies presented in Sec.~\ref{sec:sensitivity}. A more detailed description of the basic structure and features of the OSIRIS simulation framework are reported in Ref.~\cite{GensterThesis}.

\subsection{Simulation Framework}
\label{sec:simframework}

\begin{figure}[b]
	\centering
	\includegraphics[width=0.8\linewidth]{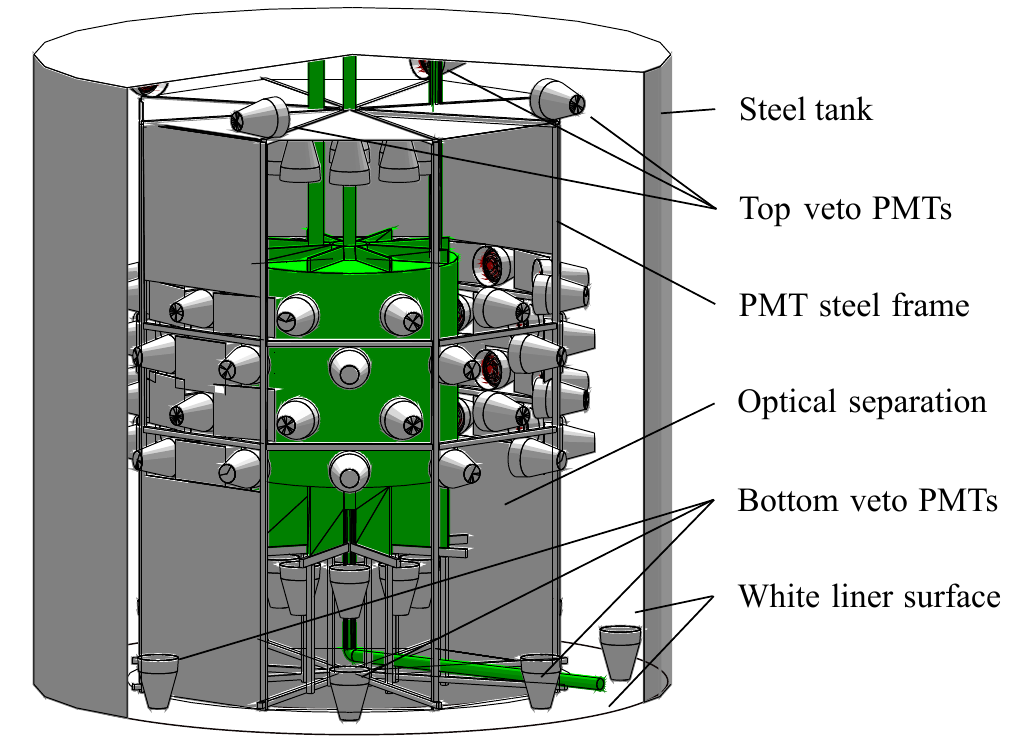} 
	\caption{Rendered depiction of the geometry implemented in simulation}
	\label{fig:simu_render} 
\end{figure}

\paragraph{Geometry implementation} The core geometry and key elements of the OSIRIS detector are implemented as illustrated in Fig.~\ref{fig:simu_render}. The basic geometry is implemented as concentric cylindrical volumes representing the LS, acrylic vessel, water shielding, steel tank and surrounding rock. The individual detector components are  constructed inside these basic volumes. Piping for the LS, calibration sources, thermorod and LS diffusers as well as the lower acrylic footing are directly attached to the AV volume. The steel frame holding the PMTs and the optical separation are inserted into the water volume. The implementation of the PMT model follows the realistic design of the Hamamatsu (R15343) 20''-PMT. The magnetic shielding is included to model its optical shading and reflectivity (Sec.~\ref{sec:pmts}). The photocathodes are defined as dedicated {\it sensitive volumes}. The photon detection efficiency of the PMTs takes into account the wavelenght depedent quantum efficiency ($\sim 28$\%) and assumed 95 \% collection efficiency.

\paragraph{Physics List} The physics list defining particles and physics processes is set up modularly and is based largely on the standard GEANT4 libraries. To model the scintillation and propagation of light signals, most of the standard interaction classes for optical photons are adopted, including the Cherenkov light creation model, photon scattering (Rayleigh and Mie) and boundary processes. Scintillation, attenuation, refractive index and reflections on surfaces and boundaries of material are implemented for all detector components. To address the characteristics of the JUNO scintillator precisely, the optical model was extended by custom definitions that address the scintillation process and the re-emission of photons with the required precision and are described in Sec.~\ref{sec:custom_simulation}.  

For modeling the interactions of particles, the definition of the detector materials includes physical properties as the density and molecular composition. The external physics list \textit{QGSP\_BERT\_HP} is initialized to simulate the production of cosmogenic isotopes by muon spallation. This provides a basis for comparison to other experiments, since this list is as well used in the JUNO simulation and has been verified by Borexino measurements \cite{Bellini:2013pxa}. 

\paragraph{Generators} To address the needs of different physics simulation studies, several primary particle generators have been implemented. In addition to the standard General Particle Source provided by GEANT4, a custom HEP mode allows to use event parameters from external generators. In order to study the external gamma background in the LS volume, a biasing mode has been implemented. The technique helps to partially overcome the need to generate zillions of gamma-rays in the surrounding rock in order to compensate the loss in statistics due to the large attenuation factor imposed by absorption in the water shield. Checking the progress of the gammas through an onion-like shell structure, the algorithm discontinues the tracking of gammas that are unlikely to deposit energy in the target volume and instead increases the weight assigned to inward-pointing tracks by creating identical copies. By correcting the final result for the modified weights, a realistic energy spectrum can be obtained with a reasonable simulation effort. The validity of the approach and the specific tuning chosen for OSIRIS have been verified against a standard simulation \cite{GensterThesis}.

In order to enable the simulation of a realistic muon flux, a dedicated event generator was established which is based on the onsite flux properties. The expected muon flux has been calculated to be  $\unit[3.7]{mHz/m^2}$ with a mean muon energy of $\unit[209]{GeV}$. This translates to an expected muon rate in the OSIRIS detector of $\unit[0.39]{Hz}$ and $\unit[0.04]{Hz}$ in the LS volume.

\subsection{Custom modules implemented for OSIRIS}
\label{sec:custom_simulation}

\begin{table}[h]
	\caption{General parameters of the LAB scintillator used in the simulation. For the wavelength-dependent properties, the value at 430 nm is quoted \cite{GensterThesis}}
	\label{tab:scint_prop}
	\centering
	\begin{tabular}{lcc}
		\toprule
		Description & Parameter & Value \\
		\midrule
		Mass density & $\rho_m$ & $0.86$ g/cm³ \\
		Light yield & $\mathcal{L}$ & 11522 MeV${}^{-1}$ \\
		Refractive index & n & 1.495 \\
		Rayleigh scattering length & $l_r$ & 30 m \\
		Absorption length & $l_{abs}$ & 85 m \\
		Re-emission probability & $p_{rem}$ & 0.30 \\
		Re-emission time constant & $\tau_{rem}$ & 1.5 ns \\
		\bottomrule
	\end{tabular}
\end{table}

\begin{table*}[ht!]
	\caption{Scintillator specific light emission properties depending on the particle type}
	\label{tab:scint_prop_particle}
	\centering
	\begin{tabular}{l|cccccc|c}
		\toprule
		\multicolumn{1}{l}{}&\multicolumn{2}{c}{Decay component 1}&\multicolumn{2}{c}{Decay component 2}&\multicolumn{2}{c}{Decay component 3} & Birk's constant \\
		Particles & $w_1$ & $\tau_1$(ns) & $w_2$ & $\tau_2$(ns) & $w_3$ & $\tau_3$(ns) &(g/cm²MeV) \\
		\midrule
		$\gamma$, e${}^-$, e${}^+$	& 0.799 & 4.93 & 0.171 & 20.6 & 0.03 & 190.0 & $1.306\cdot10^{-2}$\\
		n, p 						& 0.65 & 4.93 & 0.231 & 34.0 & 0.119 & 220.0 & $8.428\cdot10^{-3}$\\
		$\alpha$ 					& 0.65 & 4.93 & 0.228 & 35.0 & 0.122 & 220.0 & $6.106\cdot10^{-3}$ \\
		\bottomrule
	\end{tabular}
\end{table*}

\paragraph{Improved scintillation model} The original scintillation mo-del of GEANT4 describes the time profile of photon emission by a two-component model. It as well lacks a description of particle-dependent quenching. To increase the precision of the simulation for OSIRIS in both respects, we have introduced a custom-made scintillation model including three decay components for the scintillation time profile and quenching, both depending on the particle type. The general parameters describing the scintillator properties are listed in Tab.~\ref{tab:scint_prop}, while the weights and time constants describing the respective emission profiles as well as Birk's constants for different particle types are listed in Tab.~\ref{tab:scint_prop_particle}. 

In the microscopic modeling applied, the number of emitted photons for each simulation step of a charged particle is sampled from a Poisson/Normal distribution around the mean value obtained from energy deposition and light yield $\mathcal{L}$. Photons are positioned randomly along the particle track. The emission time is sampled from the particle-dependent time profile. Emission is isotropic with random linear polarization and the energy of photon sampled from the user-defined scintillation spectrum. If quenching is activated, the energy deposition and corresponding number of emitted photons is adjusted according to Birk's law.

\paragraph{Re-emission} While the propagation of scintillation photons in the LS is well-covered by the GEANT4 standard libraries, the absorption of photons by LS molecules with subsequent wavelength-shifted re-emission is not implemented. Thus, a custom class has been implemented that is called in case a photon is absorbed in the LS. The probability for re-emission is calculated as a function of wavelength. A re-emitted photon is created similarly to the custom scintillation process, with an adjusted energy spectrum and short re-emission time profile with $\tau_{\rm re}= \unit[1.5]{ns}$.

\paragraph{\ce{^9Li} and \ce{^8He} decays}
The Bi-Po coincidence signals used to determine the U/Th contamination (Sec.~\ref{sec:sensitivity}) can potentially be mimicked by the $\beta n$-decays of the cosmogenic radioisotopes \ce{^9Li} and \ce{^8He}. In GEANT4.10.2.3, this decay is not implemented correctly as the excited states of  \ce{^9Be} and \ce{^8Li} in all cases decay directly to the ground state by the emission of a gamma photon. The additional intermediate states required for a realistic simulation have been implemented in the utilized GEANT4 installation according to Ref.~\cite{Jollet2020}.

\subsection{Expected performance}
While detailed reconstruction algorithms are still under development, the studies presented here are using an effective description of the expected energy and spatial resolution based on the experience and simulation studies performed for comparable scintillation detectors (e.g.~Ref.~\cite{Bellini2014}). 
\medskip\\
The {\bf photo electron (p.e.)~yield $Y_{\rm pe}$} expected for scintillation events has been determined by a simulation of $10^5$ electrons with an energy of $E=\unit[1]{MeV}$ inserted at the detector center. For the moment, the average result of $Y_{\rm pe} = 283$\ p.e./MeV is applied independent of event position. 
\medskip\\
Based on this, the {\bf reconstructed event energy} is determined to first order by the effect of variations in p.e.~statistics. An effective description of the baseline energy resolution is obtained by applying Gaussian smearing to the true event energy $E$ with a standard deviation 
\begin{equation}
    {\sigma_{E}} = \frac{\sqrt{Y_{\rm pe} E}}{Y_{\rm pe}E} \cdot E = \sqrt{\frac{E}{Y_{\rm pe}}} \approx \unit[6]{\%}\cdot \sqrt{E/\unit[]{MeV}}
\end{equation}
The true position of a scintillation event is taken as the energy-weighted barycenter of all corresponding energy depositions inside the LS volume:
\begin{equation}
    \Vec{r}_{\rm bar} = \frac{\sum_i \epsilon_i \Vec{r}_i}{\sum_i \epsilon_i}
\end{equation}
 with the deposited energy $\epsilon_i$ at position $\Vec{r}_i$. Again, the {\bf reconstructed vertex position} is obtained by an effective description of the corresponding uncertainty. The true vertex position $ \Vec{r}_{\rm bar}$ is smeared coordinate-wise in $x$, $y$ and $z$ by sampling from a Gaussian distribution with an energy-dependent standard deviation of 
 \begin{equation}
     \sigma = \frac{\sigma_0}{\sqrt{E}} \approx \frac{\unit[14]{cm}}{\sqrt{E/\unit[]{MeV}}}
 \end{equation}
 with the true event energy $E$. The value of $\sigma_{0}=\unit[14]{cm}$ is scaled from the spatial resolution achieved in the Borexino experiments \cite{Bellini2014} taking into account the difference in $Y_{\rm pe}$ between the detectors. The corresponding three-dimensional position uncertainty is $\sim$\unit[24]{cm}.
\medskip\\
A preliminary {\bf physics trigger threshold} $E_{\rm thr}$ has been determined based on a simulation of the signals from low-energy \ce{^{14}C} decays that will dominate the low-energy end of the OSIRIS event spectrum and the expected PMT dark noise levels. For this, a total of $2.5 \cdot 10^6$ \ce{^{14}C} events has been simulated with a uniform distribution in the LS volume. In the current implementation, the physics trigger of the EventBuilder considers the PMT hit multiplicity within a given time window (Sec.~\ref{sec:daq}). So-called ''dark triggers'' may occur based on accidental coincidences of uncorrelated dark counts from several PMTs within the time window. The study shows that an acceptable dark trigger rate ($<$1\,Hz) can be achieved in case the trigger condition is set to 5 PMT hits within $\unit[70]{ns}$. This corresponds to a trigger efficiency of $\unit[90]{\%}$ at a nominal trigger threshold of $E_{\rm thr}\approx\unit[36]{keV}$.

\subsection{Background Levels}
The simulation framework has been used to predict the background levels from natural and cosmogenic radioactivity that will impact the OSIRIS physics analyses. For the primary objective, i.e.~the determination of the U/Th background levels from detecting Bi-Po coincidence events, accidental coincidences from single-event background events play an important role. Simulations have helped to determine the required size of the water buffer surrounding the LS volume, the distance of the inner PMT array from the AV and to set radiopurity limits for all internal components of the OSIRIS detector (Sec.~\ref{sec:radioactivity}). Moreover, selection cuts for Bi-Po coincidence events have been devised and optimized based on simulated Bi-Po signal and background spectra (Sec.~\ref{sec:bipo_cuts}). Finally, the time-correlated occurrence of cosmic muons and their secondary spallation products has been investigated to derive the corresponding correlated event rates and set minimum requirements for the muon veto efficiency (sections \ref{sec:muonveto} and \ref{sec:cosmogenic}).

\subsubsection{Natural Radioactivity}
\label{sec:radioactivity}
Most radioactive decays produce single events inside the detectors and will only emerge as a background to the Bi-Po coincidence search in case their level is sufficiently high to create a large rate of accidental coincidences. Due to the stringent radiopurity requirements set for the JUNO LS (see Sec.~\ref{sec:sensitivity}), radioactive decays inside the LS volume of OSIRIS will be relatively rare compared to the external background level \cite{An:2015jdp}. 

\begin{figure}[h!]
	\centering
	\includegraphics[width=0.9\linewidth]{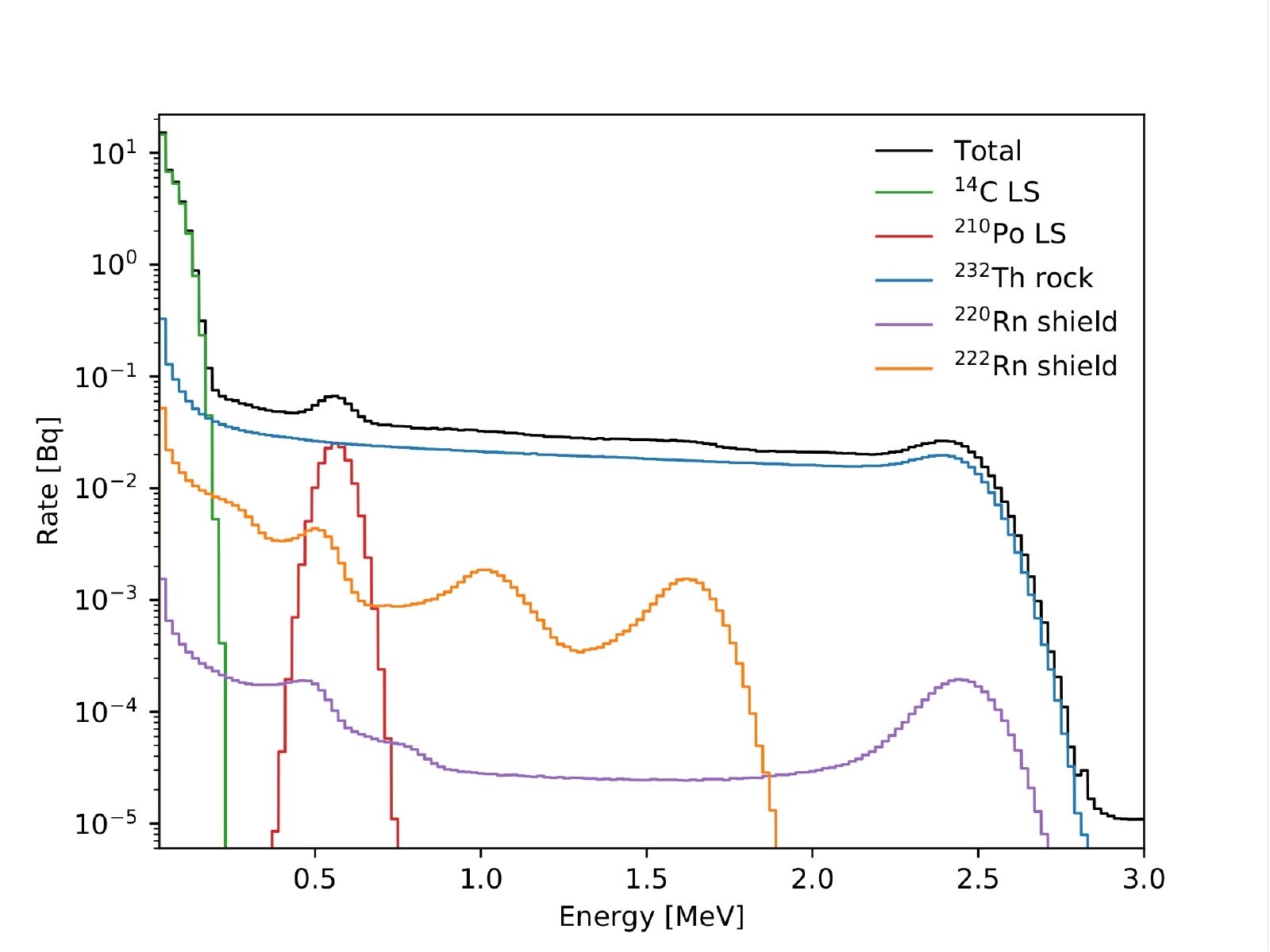} 
	\caption{Energy spectrum of the total background rate expected in the full LS volume including Radon contamination of the water shield. Only the contributions that define the shape are overlayed to show their influence in the different energy regions (based on \cite{GensterThesis})}
	\label{fig:single_bg_spec} 
\end{figure}

\paragraph{External background} We find that gamma-rays emitted in the decays of natural radioactivity in the outer detector materials and surrounding rock are dominating the single-event spectrum. The main isotopes to consider are \ce{^{40}K} and the elements of the \ce{^{238}U} and \ce{^{232}Th} decay chains. A corresponding study has been performed based on the biasing algorithm described in Sec.~\ref{sec:simframework}. The resulting event rates and associated energy spectrum of the main contributors are shown in table \ref{tab:int_gamma} and figure \ref{fig:single_bg_spec}. Despite 3\,m of water shielding, the dominant background source are elements of the \ce{^{232}Th} chain (especially \ce{^{208}Tl}) in the surrounding rock, followed by the contribution of decays in the PMT glass.

\paragraph{Radon in the water shield} The possible impact of radon dissolved in the water surrounding the AV has been investigated. Given the low radon content ($\leq$1\,mBq/m$^3$) of the JUNO high-purity water that is used as a supply for the WT, the main contribution of radon inside the water shield is expected from the emanation of detector materials. To study the impact, we conservatively assume full intermixing and thus a uniform distribution of radon in the water volume. Based on the U/Th abundances of the materials and emanation rates measured for JUNO and other low-background experiments, we expect final concentrations of $\sim$70\,mBq/m$^3$ for \ce{^{222}Rn} and $\sim$1.4\,mBq/m$^3$ for \ce{^{220}Rn} in water. The rates are dominated by emanation from the PMT glass and potentially cables. All radon daughters have been simulated and the emitted $\gamma$-rays propagated into the LS volume to obtain the spectra shown along with the other external backgrounds in Fig.~\ref{fig:single_bg_spec}. At the expected emanation levels, the radon contribution is subdominant.

\begin{table}[h!]
	\caption{The main contributors to the single-event background in OSIRIS. The external background from $\gamma$-emission in the surrounding rock and from PMT glass are dominating compared to a potential radon contamination of the water shield (based on \cite{GensterThesis})}
	\label{tab:int_gamma}
	\centering
	\begin{tabular}{lccc}
		\toprule
		Component & Isotope & Contamination (Bq/kg) & Rate in LS($s^{-1}$) \\
		\midrule
		\multirow{3}{*}{Rock} & ${}^{40}$K & 220 & $1.3\times10^{-2}$ \\
		& ${}^{232}$Th & 123 & 3.5 \\
		& ${}^{238}$U & 142 & $2.8\times10^{-1}$ \\
		\midrule
		\multirow{3}{*}{PMT glass} & ${}^{40}$K & 1.89 & $2.0\times10^{-2}$ \\
		& ${}^{232}$Th & 1.72 & $9.0\times10^{-1}$ \\
		& ${}^{238}$U & 4.77 & $6.1\times10^{-1}$ \\
		\midrule
		\multirow{2}{*}{Water shield} & ${}^{220}$Rn & $1.4\times10^{-6}$ & $1.8\times10^{-2}$ \\
		& ${}^{222}$Rn & $6.8\times10^{-5}$ & $4.2\times10^{-1}$ \\
		\bottomrule
	\end{tabular}
\end{table}

\subsubsection{Cosmogenic Background}
\label{sec:cosmogenic}

The expected background contribution induced by cosmic muons was determined based on a simulation sample of $10^8$ muons crossing the OSIRIS detector volume (including the \textit{QGSP\_BERT\_HP} physics list, Sec.~\ref{sec:simframework}). In the generation of the muons, the angular and energy distributions expected for the underground Scintillator Hall were taken into account. With regards to the Bi-Po coincidence search, the relevant events are generated by secondary particles from muon spallations, especially neutrons and the $\beta n$-emitting radioisotopes \ce{^8He} and \ce{^9Li}. Since their generation and capture/decay are correlated in time, there chance to mimic a fast Bi-Po-coincidence is considerably enhanced compared to single-event background.

\paragraph{Two-neutron events} The largest contribution emerges from the accidental coincidences of neutrons produced by the same primary cosmic muon. About 1\% of the simulated muons produced at least one neutron to be captured on hydrogen in the LS volume, releasing a 2.2\,MeV gamma-ray after a typical capture time of $\tau_n\sim250$\,\textmu{}s. In order to mimic a Bi-Po coincidence, the gamma event of the second neutron capture must be only partially contained within the LS volume in order to fall into the energy range selected for Po decays (Sec.~\ref{sec:bipo_cuts}). The expected rate after the application of the Bi-Po selection cuts is $\unit[(0.125\pm0.008)]{d}^{-1}$ in the LS volume. 

\paragraph{$\beta n$-emitters} The prompt $\beta$-decay of \ce{^8He} and \ce{^9Li}, followed by the emission of a neutron from \ce{^8Li} and \ce{^9Be} that is subsequently captured on hydrogen can potentially mimic the fast coincidence signature. However, also in this case the $n$-capture gamma must only partially deposit its energy in the LS volume to pass the energy cut for the delayed $\alpha$-decay. Hence, the expected event rate after Bi-Po selection cuts is only $\sim \unit[(7.4\pm 0.6)\cdot 10^{-7}] {d}^{-1}$.

\paragraph{Rock neutrons} The two-neutron event rate can be further reduced by the application of a veto for which the muon and all subsequent events in the following milliseconds are excluded from analysis. However, neutrons may be generated as well from muons passing through the surrounding rock, in which case the parent muon cannot be detected. We analyzed the corresponding neutron rate by increasing the simulation volume to a cube of $\sim$\unit[10]{m} to permit neutron production in the cavern rock. The expected rate of fast neutrons, arriving at the LS volume is 4.7 per day, resulting in $\sim$4 captures per day. The contribution of fast neutrons to the coincident background, in combination with other neutrons or gamma photons, is negligible \cite{bieger_thesis}.

\section{Sensitivity}
\label{sec:sensitivity}


Trace contaminations of radioactive isotopes in the LS represent a major background for the JUNO neutrino program. While the JUNO LS system has been devised to purify the LS to an acceptable radioactivity level, it is the main purpose of the OSIRIS detector to verify the required quality is achieved. The corresponding requirement levels are lined out in Table~\ref{tab:radiopurity}. Upper limits have been set for the acceptable activity of elements in the natural uranium and thorium chain as well as individual limits for \ce{^{210}Po}, \ce{^{40}K} and \ce{^{14}C}. For comparison, we give the radiopurity levels that have been achieved in the KamLAND and Borexino experiments \cite{Gando:2014wjd,Agostini:2018uly}. 



\begin{table*}[h]
\caption{Baseline radiopurity requirements set for JUNO in comparison to the experimental values achieved in KamLAND and Borexino \cite{Gando:2014wjd}\cite{Agostini:2018uly}\cite{Keefer:2011kf}}
\label{tab:radiopurity}
\centering
\begin{tabular}{l|cc|cc}
\toprule
& JUNO IBD & JUNO solar & KamLAND & Borexino \\
Chain/Isotope & [g/g] & [g/g] & [g/g] & [g/g] \\
\midrule
$^{238}$U & $1\times 10^{-15}$ & $1\times 10^{-16}$ & $(5.0\pm0.2)\times 10^{-18}$ & $<1\times 10^{-18}$\\
$^{232}$Th & $1\times 10^{-15}$ & $1\times 10^{-16}$ & $(1.3\pm0.1)\times 10^{-17}$ & $<1\times 10^{-18}$\\
\midrule
$^{210}$Po & $-$ & $5\times 10^{-24}$ & $\sim 2 \times 10^{-23}$ & $<1\times 10^{-25}$\\
$^{40}$K & $1\times 10^{-16}$ & $1\times 10^{-17}$ & $(7.3\pm1.2)\times 10^{-17}$ & $<1\times 10^{-19}$\\
$^{14}$C & $1\times 10^{-17}$ & $1\times 10^{-17}$ & $(3.98 \pm 0.94) \times 10^{-18}$ & $(2.7\pm0.1)\times 10^{-18}$\\
\bottomrule
\end{tabular}
\end{table*}

The OSIRIS design has been optimized for sensitivity for the U/Th decay rates via the tagging of fast Bi-Po coincidence decays towards the lower end of the U/Th decay chains. Corresponding sensitivities in different operation modes of OSIRIS are discussed in Sec.~\ref{sec:bipo}. Moreover, OSIRIS will be able to monitor the levels of $^{14}$C and $^{210}$Po since they provide distinctive spectral features that can be identified over the external gamma-ray background (Sec.~\ref{sec:singles}).

\begin{table*}[h]
\caption{ Basic parameters of Bi-Po coincidence decays and their link to U/Th concentrations.}
\label{tab:bipo}
\centering
\begin{tabular}{cc|cc|c}
\toprule
Decay chain & Mass limit (IBDs) & $\beta$-$\alpha$-coincidence & Activity limit & Po lifetime \\
\midrule
$^{238}$U & $10^{-15}$\,g/g & $^{214}$Bi-$^{214}$Po & $1.2 \times 10^{-8}$\,Bq/kg & 237\,\textmu s\\
$^{232}$Th & $10^{-15}$\,g/g & $^{212}$Bi-$^{212}$Po & $0.4 \times 10^{-8}$\,Bq/kg & 0.5\,\textmu s\\
\bottomrule
\end{tabular}
\end{table*}

\subsection{Uranium and thorium chains}
\label{sec:bipo}

In a LS detector, the most sensitive direct method to determine the abundances of \ce{^{238}U} and \ce{^{232}Th} in the LS is the tagging of the fast coincidences of $^{214}$Bi-Po and $^{212}$Bi-Po, respectively. Governed by the short lifetimes of the polonium isotopes ($\sim$\textmu{}s), the double-event signature offers a potent possibility to reject single-event backgrounds. Assuming secular equilibrium, the measured Bi-Po rates can be directly translated to a mass abundance of U/Th. The corresponding relations are summarized in Tab.~\ref{tab:bipo}.

In this section, we firstly discuss the event selection to isolate true Bi-Po coincidences from accidental backgrounds (Sec.~\ref{sec:bipo_cuts}). We then regard the most simple case in which the assumption of secular equilibrium within the U/Th chains is true and OSIRIS can reach its best sensitivity (Sec.~\ref{sec:basic_sensitivity}). However, it is clear that the Bi-Po search in OSIRIS will be beset by a background created by radon emanation in the JUNO LS system. The corresponding sensitivities in {\it batch-mode} and {\it continuous-mode} operations are discussed in the last sections \ref{sec:bipo_batch} and \ref{sec:bipo_cont}.

\subsubsection{Bi-Po Selection Cuts} \label{sec:sensitivity:BiPo:stat}
\label{sec:bipo_cuts}

\begin{figure}[t]
    \centering
    \includegraphics[width=0.45\textwidth]{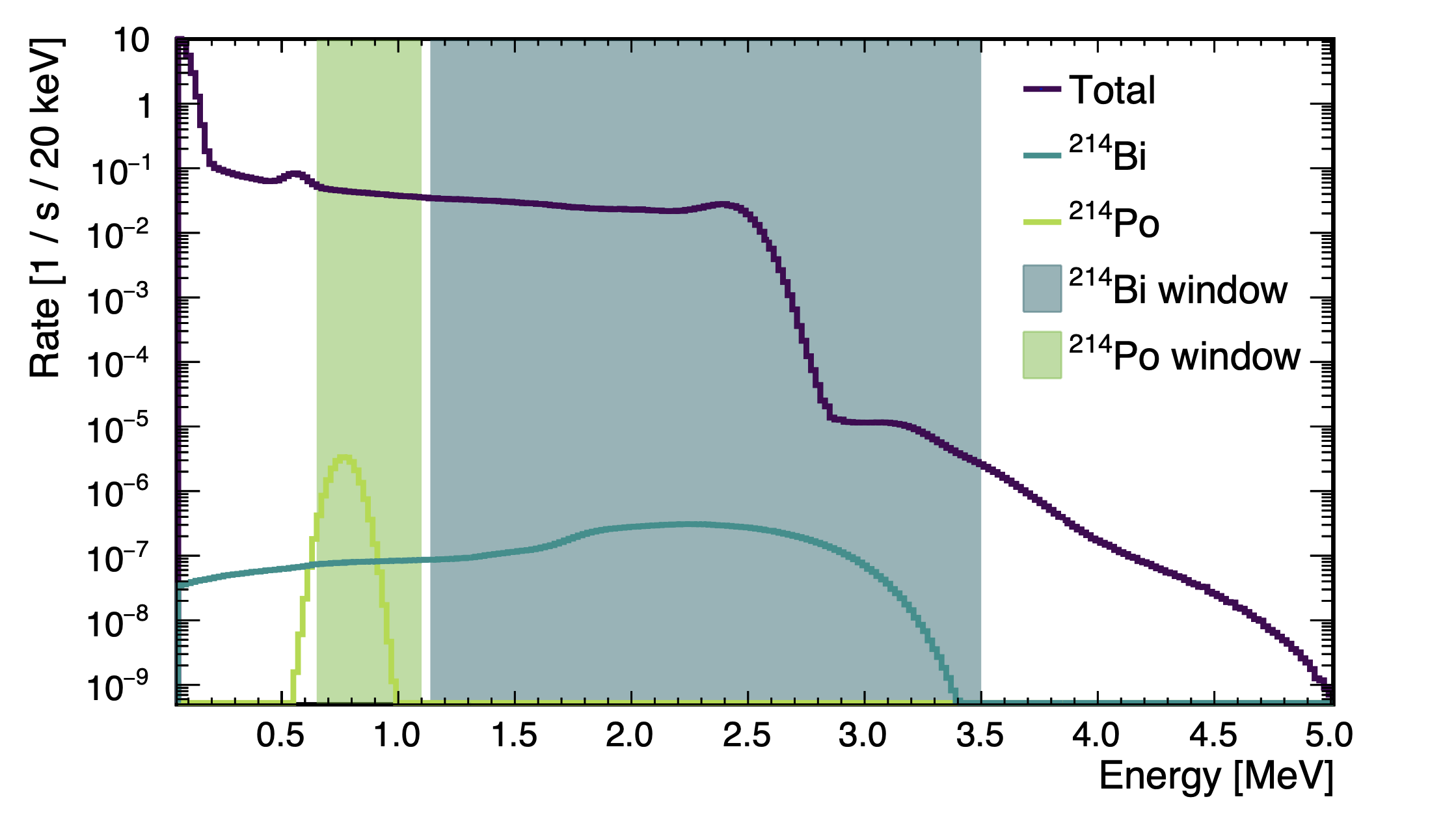}
    \caption{Bi-Po coincidences are found based on a sequence of selection cuts. Here, we display the energy selection cuts for prompt and delayed events as well as the underlying signal and background spectra for $^{214}$Bi-Po}
    \label{fig:bipo_selection}
\end{figure}

The selection of Bi-Po coincidence events relies on a sequence of four consecutive cuts:

\begin{itemize}
    \item {\bf Fiducial Volume (FV).} While the external gamma-ray background is substantially reduced by the water shielding, the rate of external events close to the verge of the AV is enhanced compared to the bulk volume of the LS. Moreover, we expect a contribution from residual contaminations on the AV surface. Therefore, spatial reconstruction is used to define a cylindrical FV cut of 2.8\,m in height and diameter for the prompt event.
    
    \item{\bf Energy selection.} The $\beta$-spectra of bismuth feature a rather broad energy spectrum, while the $\alpha$-decays of polonium are mono-energetic of a given isotope. As presented in Fig.~\ref{fig:bipo_selection}, the visible signal is shifted to the range of $0.5-1$\,MeV due to quenching. In the following, the term ''Bi-candidate'' describes all events which are inside of the Bi selection cut, and likewise for Po events. 
    
    \item{\bf Timing cut.} The third and most powerful cut to reject accidental coincidences is the time-delay cut. Given the very short lifetimes of \SI{431}{\nano\second} and \SI{237}{\micro\second} for $^{212}$Po and $^{214}$Po, respectively, most single-event backgrounds can be rejected by requiring a short time delay $\Delta t$ for the detection of the subsequent Po-candidate. We conservatively assume that a minimum $\Delta t$ of  \SI{200}{\nano\second} is required to distinguish prompt and delayed event \cite{GensterThesis}. The upper bound is adjusted to optimize sensitivity. 
    
    \item{\bf Distance Cut.} Given the low recoil energies and short $\Delta t$, both Bi and Po decays will occur at virtually identical positions, smeared out by position reconstruction. Accidental coincidences, conversely, can be expected to be more-or-less randomly distributed over the entire FV. Imposing a maximum value for the distance  $\Delta r$ between Po and Bi-candidates efficiently rejects accidental backgrounds while keeping most of the signal.
\end{itemize}

\noindent The corresponding parameter ranges and efficiencies of the selection cuts are listed in Table~\ref{tab:stat:efficiencies}. Assuming a contamination level of $10^{-15}$\,g/g for uranium (thorium) and applying the event selection cuts, we expect to identify $\dot n_{\rm Bi-Po}=11$ (4) coincidence events per day for $^{214}$Bi-Po ($^{212}$Bi-Po), while the corresponding residual background is $\dot n_{\rm acc}=1.14$ (0.04) accidental coincidences per day. 

\begin{table*}[h]
    \caption{Efficiencies of the different data selection cuts (see text). The fiducial volume is defined as a cylinder of 2.8\,m height and diameter, $E_{\rm Bi,Po}$ are the energy windows selected for prompt and delayed decays, $\Delta t$ and $\Delta r$ are time and distance cuts. For the latter, the numbers in brackets represent the rejection efficiency for accidental coincidences.}
    \label{tab:stat:efficiencies}
    \centering
    \begin{tabular}{c | c c | c c}
        \toprule
         &  \multicolumn{2}{c|}{\ce{^{212}Bi-Po}} & \multicolumn{2}{c}{\ce{^{214}Bi-Po}}\\
        \midrule
        {} Cut  & Range   & Efficiencies (\%)    & Values   & Efficiencies (\%)\\
        \midrule
        FV & $<$\SI{140}{\centi\meter} & - & $<$\SI{140}{\centi\meter} & - \\
        $E_{\rm Bi}$   &  \SIrange{0.14}{2.5}{\mega\electronvolt} & 98   & \SIrange{1.4}{3.5}{\mega\electronvolt}  & 83\\
        $E_{\rm Po}$  &  \SIrange{0.9}{1.5}{\mega\electronvolt} & 100   & \SIrange{0.7}{1.1}{\mega\electronvolt}  & 99\\
        $\Delta t$  &  \SIrange{0.2}{2}{\micro\second}  &  62   & \SIrange{0.2}{711}{\micro\second}  & 95\\
        $\Delta r$ & $<$\SI{72.5}{\centi\meter} & 98 (94) & $<$\SI{45}{\centi\meter} & 88 (98)\\
        \midrule
        Total &  & 60 & & 69\\
        \bottomrule
    \end{tabular}
\end{table*}


\subsubsection{Basic Statistical Analysis}
\label{sec:basic_sensitivity}

\begin{figure}[t]
    \centering
    \includegraphics[width=0.45\textwidth]{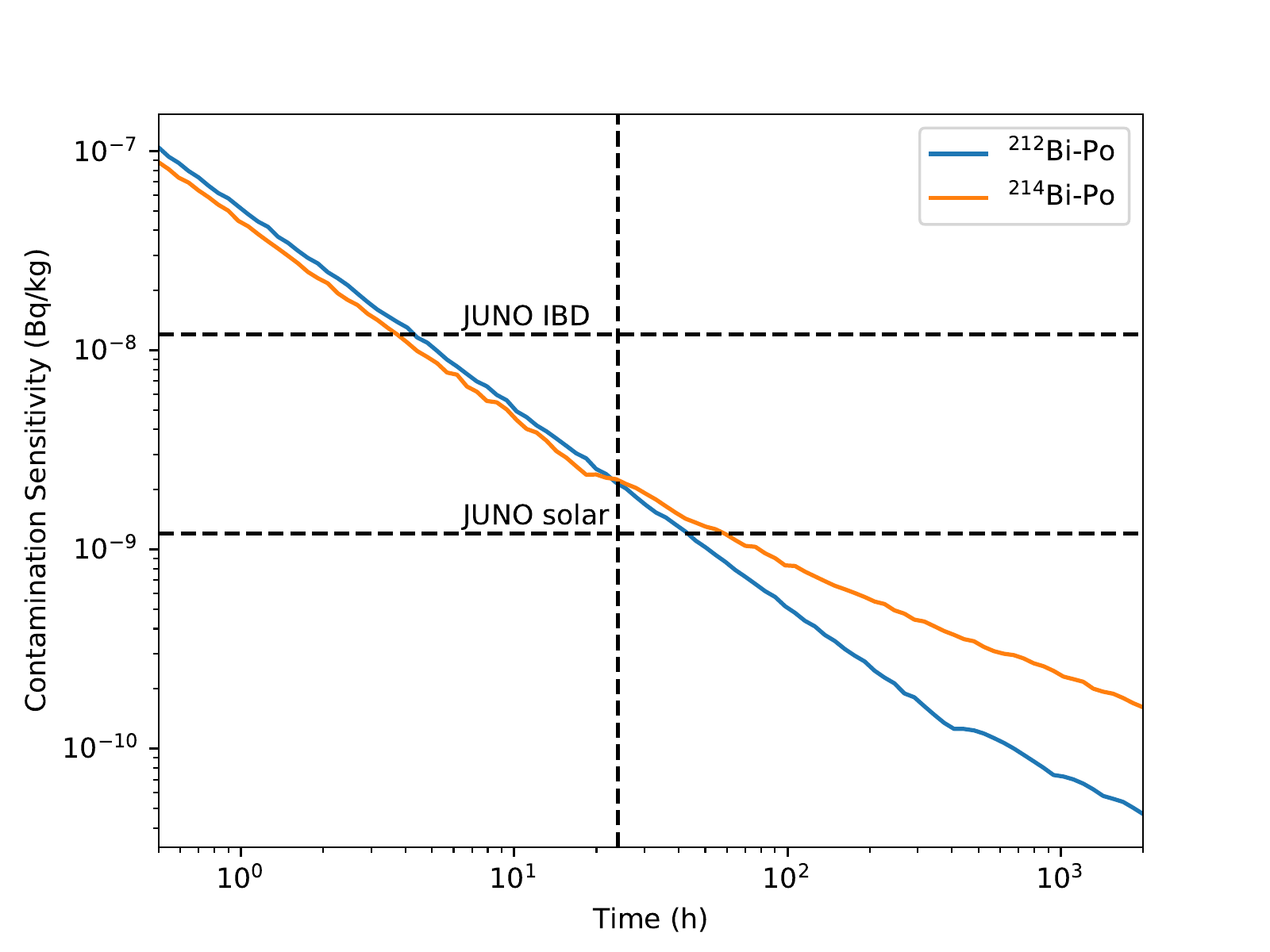}
    \caption{Sensitivity of OSIRIS as a function of the measurement time for the $^{212}$Bi-Po (blue) and $^{214}$Bi-Po (orange) coincidence searches. The fluctuations in the curves reflect the innate uncertainty of the Monte-Carlo method (see text). Within a few hours, OSIRIS can verify the JUNO IBD requirement. In order to confirm the contamination level of the JUNO solar requirement, a measurement longer than one day is needed}
    \label{fig:sensitivity}
\end{figure}

The sensitivity of OSIRIS to a radon contamination in the LS is defined as the signal rate of Bi-Po decays for which the selected coincidence rate $\dot n_{\rm BiPo}$ exceeds the level of background fluctuation from accidental coincidences $\dot n_{\rm acc}$ on a 90$\%$ confidence level (CL). This sensitivity depends on the absolute number of detected events
\begin{equation}
    n(T) = (\dot n_{\rm BiPo} + \dot n_{\rm acc})\cdot T
\end{equation}
and thus on the measuring time $T$ \cite{GensterThesis}. For short $T={\cal O}(1{\rm d})$, the expected low values of $n(T)$ forbid the use of Gaussian approximations and necessitate a more careful calculation of the corresponding upper limits on $\dot n_{\rm BiPo}$. Therefore, we adopt the strategy for sensitivity determination of a counting experiment as described in Ref.~\cite{Cowan}. The corresponding sensitivity curves for a 90\,\% confidence level are shown in Fig.~\ref{fig:sensitivity}. 

\noindent In the most basic scenario, we assume that the U/Th decay chains are in secular equilibrium. This means that the limit on the Bi-Po rate $\dot n_{\rm BiPo}$ can be directly interpreted as a limit on the corresponding U/Th decay rate $\dot n_{\rm U/Th}$ that can be converted to a mass limit as illustrated by Table \ref{tab:bipo}. Based on this, {\it IBD-level} radiopurity (i.e. $10^{-15}$\,g/g of U/Th) can be tested within several hours. Instead, a few days are needed to approach the {\it solar level} of $10^{-16}$\,g/g. The U sensitivity is slightly better for short times $T$ because of the higher selection efficiency. However, the Th analysis remains for a longer time in the background-free regime. Thus, the sensitivity surpasses that of U for $T>1$\,d. 

\subsubsection{Batch-mode operation}
\label{sec:bipo_batch}

In the basic analysis, we have neglected the presence of correlated backgrounds, most noteworthy a radon contamination in the LS that is not in secular equilibrium with the U/Th decay chain. The relevant radon isotopes, i.e.~\ce{^{222}Rn} for \ce{^{238}U} and \ce{^{220}Rn} for \ce{^{232}Th}, are relatively short-lived. So, they do not pose a threat to the JUNO physics program. However, their short-term decays represents an additional source of Bi-Po coincidences for OSIRIS that is superimposed to U/Th-supported Bi-Po rate. Correct treatment of this off-equilibrium component is crucial to obtain a reliable U/Th contamination estimate.

Radon will emanate from the surfaces of all pipes and vessels coming into contact with LS on its way through the JUNO filling system. Moreover, gas leakage at valves and connections will add a low amount of radon from the ambient air. From estimates based on the system surfaces, emanation and leakage rates, the radon contamination expected during JUNO filling is on the order of 7--45 decays per day and ton, corresponding to an initial rate of $\dot n_{\rm Rn}(0) = 130-830$ counts per day in the OSIRIS LS volume. This is 1--2 orders of magnitude larger than the aimed for U/Th induced decay rates, thus severely affecting short-term sensitivity.

A possibility to overcome this problem for a given LS sample is an extended measurement campaign with $T\geq\tau_i$ that provides for a visible decay of the radon component. The relevant decay times are $\tau_{222}=5.5$\,d of \ce{^{222}Rn} in the case of the U chain and $\tau_{220}=15$\,h of \ce{^{212}Pb} for the Th chain. We assume that at the start of such a {\it batch-mode} measurement, the total LS within the AV volume is replaced with a fresh batch of LS on a relatively short time scale of less than 1 day. This batch of LS is then kept inside the AV for a period $T$ of several days or weeks. The Bi-Po rate induced by radon, $\dot n_{\rm Rn}(t)=\dot n_{\rm Rn}(0)\exp(-t/\tau_i)$, will decay over time, while the U/Th-fed rate $\dot n_{\rm U,Th}$ will remain constant. 

In principle, the sensitivity will slowly improve with $T$ when applying a simple counting analysis as described in Sec.~\ref{sec:basic_sensitivity} based on the improving ratio of $n_{\rm U,Th}(T)$ to $n_{\rm Rn}(T)$. Here, however, we apply a different approach that exploits the known time behavior of the radon background. This can be achieved by a time-dependent fit to the observed coincidence rate 
\begin{equation}
\dot n_{\rm BiPo}(t)= \dot n_{\rm U/Th} + \dot n_{\rm Rn}(0)\cdot \exp\left(-\frac{t}{\tau_{220/2}}\right).
\label{eq:time_fit}
\end{equation}
To determine the median sensitivity, we assume an Asimov data set with an initial Rn rate $\dot n_{\rm Rn}(0)$ of 100 (300) counts per day and both a constant U/Th-supported rate $\dot n_{\rm U,Th}\sim 0$ and an accidental background rate set to the expectation value $\dot n_{\rm acc}$. $\dot n_{\rm U/Th}$ and $\dot n_{\rm Rn}(0)$ are the free fit parameters. As expected, we find the median result for $\dot n_{\rm U,Th}$ compatible with 0, and use the corresponding fit uncertainty to derive a (one-sided) upper limit.

\begin{figure}[t]
 \centering 
 \includegraphics[width=0.45\textwidth]{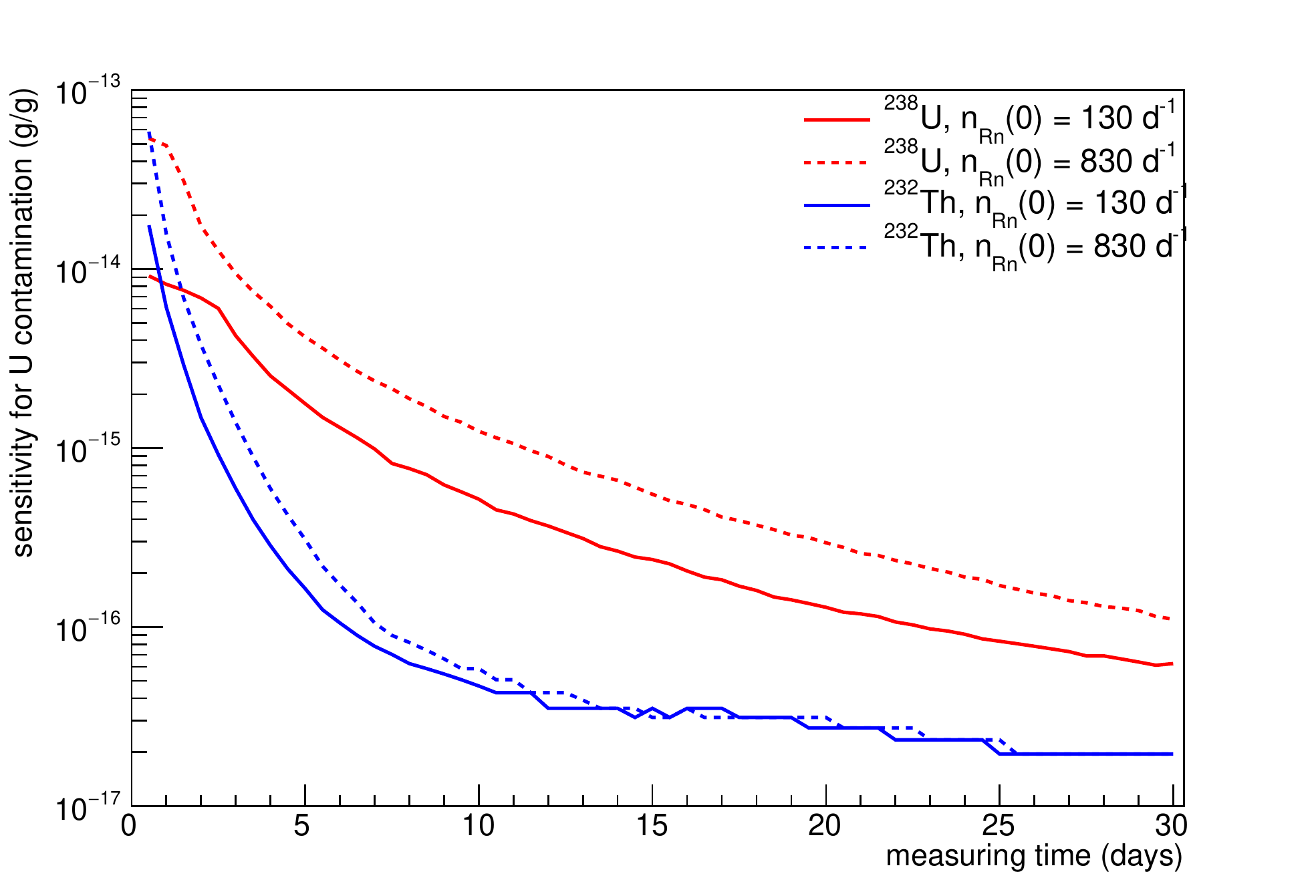}
 \caption{Sensitivity to U/Th in the presence of an initial radon contamination $Rn(0)$. An upper limit on the constant U/Th-supported term is derived from a time fit to the observed Bi-Po-rate, taking the known decay profiles of \ce{^{222}Rn}/\ce{^{220}Rn} into account}
 \label{fig:sensitivity_rn}
 \end{figure}

Figure~\ref{fig:sensitivity_rn} shows the resulting U/Th sensitivity (90\ \% C.L.) as a function of the measuring time $T$. For $\dot n_{\rm Rn}(0) = 130$, setting an upper limit on the U (Th) rate corresponding to the {\it IBD level} will take 7 (2.5) days, while for {\it solar-level} sensitivity 22 (6) days are required. 

\begin{figure*}[t]
    \centering
    \includegraphics[width=\textwidth]{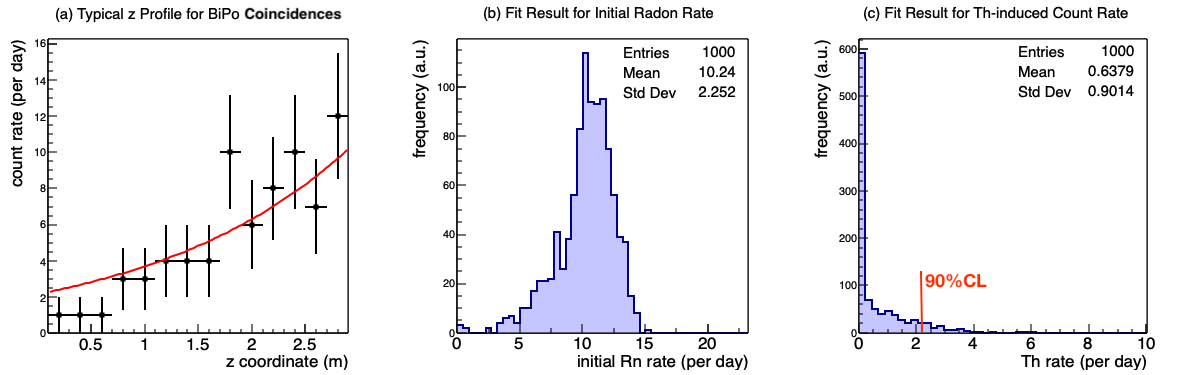}
    \caption{Th-rate determination in {\it continuous-mode} operation, assuming 830 initial Rn counts per day and AV volume. The time-dependence of the \ce{^{220}Rn} decay rate is mapped on the reconstructed Bi-Po vertex height in the AV. This height-dependence is used to fit a decaying \ce{^{220}Rn} and a $z$-constant Th-induced \ce{^{212}Bi-Po} decay rate}
    \label{fig:zdependence}
\end{figure*}
 
\subsubsection{Continuous operation}
\label{sec:bipo_cont}

While we expect that measuring times of several weeks for a single LS batch can be realized during the commissioning phase, this is not practical during the JUNO filling phase. Instead, we foresee a {\it continuous-mode} measurement, where fresh LS will be constantly fed from the top into the AV while already screened LS will be drained from the AV bottom at the same rate (Sec.~\ref{sec:lhs}). Since at a flow rate of 1\,m$^3$/h the LS passage time through the AV amounts to almost a day, the secondary Bi-Po rate supported by radon will partially decay while travelling from top to bottom of the AV. A reduction by 17\% (80\%) is expected for \ce{^{222}Rn} (\ce{^{220}Rn}). 

This decay pattern can be used to distinguish the height-dependent radon-induced Bi-Po-rate $\dot n_{\rm Rn}(z)$ from the $z$-inde-pendent U/Th-fed coincidence rate $\dot n_{\rm U/Th}$. The underlying analysis method is quite similar to that in batch-mode operation (Sec.~\ref{sec:bipo_batch}). Assuming a time-constant flow speed of individual LS layers in the AV, the reconstructed vertex height $z$ of Bi-Po coincidences can be mapped equivalently to a time coordinate $t$. In Fig.~\ref{fig:zdependence}(a), the fiducial volume is split into 14 layers of 20\,cm thickness to form a height histogram of the \ce{^{212}Bi-Po} count rate. By averaging over the Bi-Po rates registered within a day in those layers and then fitting with an effective exponential decay (plus constant) that reflects the \ce{^{212}Pb} life time and LS flow speed this can be translated in a time-dependent decay. Panels (b) and (c) of Fig.~\ref{fig:zdependence} show the results for Rn and Th-induced Bi-Po rates per slice assuming $\dot n_{\rm Rn}=830\,{\rm d}^{-1}$ and $\dot n_{\rm U/Th}\sim0$. The expected corresponding upper limit on the \ce{^{232}Th} contamination for 90\% C.L.~translates to $1.4\cdot10^{-14}$\ g/g level, or $6\cdot10^{-15}$\ g/g if $\dot n_{\rm Rn}=130\,{\rm d}^{-1}$.

The situation can be somewhat improved if the initial radon inflow is time-stable enough to provide a solid prediction of the amount of off-equilibrium radon present in the LS. If the time fit of Equation~(\ref{eq:time_fit}) is further refined to include a penalty term for the initial radon rate at the 100\% level (i.e.\ assuming that the radon rate does not vary by more than a factor 2 from day to day),  the corresponding sensitivity improves to a few times $10^{-15}$\ g/g for both U/Th. If the prior can be lowered to 10\%, e.g.~based on dedicated long-term batch measurements, the limit would come close to $10^{-15}$\ g/g. This means that even in continuous mode, OSIRIS can maintain a U/Th sensitivity close to the {\it IBD level}. 

\begin{table*}[t!]
\centering
\caption{ U/Th sensitivities expected in continuous operation mode for a 24\,h measurement, taking into account the initial presence of radon in the LS. The analysis uses the $z$-dependence of the detected Bi-Po rate introduced by radon decay. While some prior knowledge on the injected \ce{^{222}Rn} contamination is required to reach U sensitivity close to the IBD level, \ce{^{220}Rn} decays sufficiently fast to show a distinct height gradient, permitting a high-quality limit on the Th contamination}
\label{tab:radon}
\begin{tabular}{ll|cc|cc}
\toprule
 & & \multicolumn{2}{|c}{Limit on \ce{^{238}U}} & \multicolumn{2}{|c}{Limit on \ce{^{232}Th}} \\
\midrule
\multicolumn{2}{l|}{Initial Rn rate} & 130\,d$^{-1}$ & 830\,d$^{-1}$ & 130\,d$^{-1}$ & 830\,d$^{-1}$ \\
\midrule
Rn prior: & none & $7\cdot 10^{-15}$\ g/g & $4\cdot 10^{-14}$\ g/g & $6\cdot 10^{-15}$\ g/g & $1.4\cdot 10^{-14}$\ g/g\\
& 100\% & $3\cdot 10^{-15}$\ g/g & $2.6\cdot 10^{-14}$\ g/g & $5\cdot 10^{-15}$\ g/g & $1.4\cdot 10^{-14}$\ g/g\\
& 10\% & $1.0\cdot 10^{-15}$\ g/g & $3\cdot 10^{-15}$\ g/g & $3\cdot 10^{-15}$\ g/g & $9\cdot 10^{-15}$\ g/g \\
\bottomrule
\end{tabular}
\end{table*}

\subsection{\ce{^{14}C} and \ce{^{210}Po} Contaminations}
\label{sec:singles}

\begin{figure}[!hb]
\centering
\renewcommand\figurename{Figure}
\includegraphics[width=0.48\textwidth]{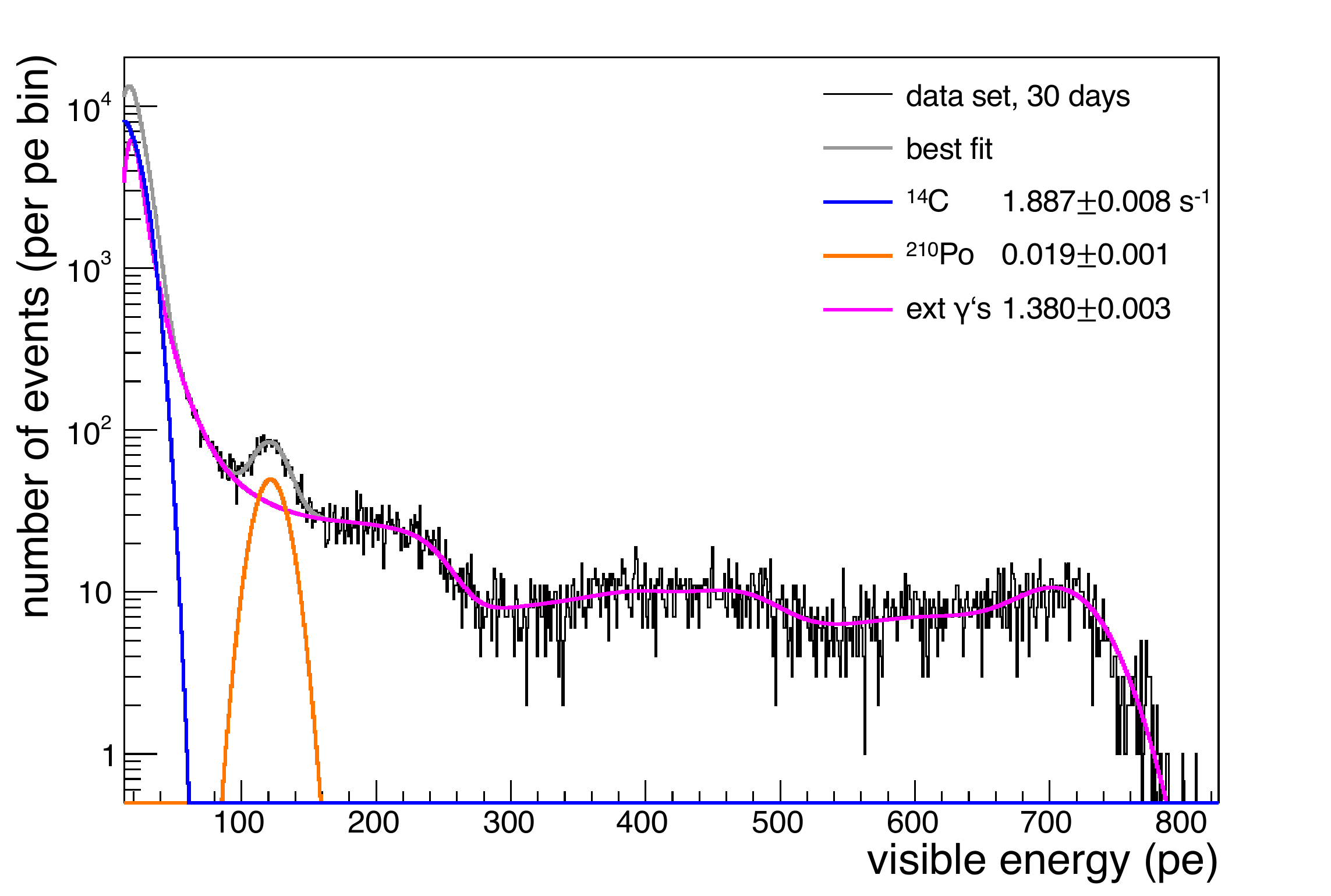}   
\caption{Sample photo electron (p.e.) spectrum for OSIRIS, applying a fiducial volume cut to the innermost cylindrical volume of 1\,m height and diameter for internal signals. The spectral contributions of $^{14}$C and $^{210}$Po emerge above the otherwise dominant external $\gamma$ background level. Corresponding rates can therefore be extracted via a spectral fit at \%-level precision. In this plot, we assume a $^{14}$C abundance of $10^{-18}$. \ce{^{210}Po} is set to the minimum requirement defined for JUNO solar neutrino analysis, i.e.~80 decay per day and ton. The external background level has been determined by simulation (sec.~\ref{sec:simulation})}
\label{fig:c14po210}
\end{figure}

Unlike in the case of the U/Th chain isotopes, the contamination levels of the LS for \ce{^{14}C} and \ce{^{210}Po} have to be determined based on the single event spectrum. A corresponding event spectrum is shown in Figure~\ref{fig:c14po210}, assuming a \ce{^{14}C} rate corresponding to a relative abundance of $10^{-18}$ and the \ce{^{210}Po} rate specified as baseline requirement for the JUNO solar neutrino program. For illustration we choose here a relatively long measuring period (30 days) while restricting the fiducial volume to a central cylinder of 1\,m height and diameter. The spectral features generated by \ce{^{14}C} and \ce{^{210}Po} are thus easily visible. 

\paragraph{\ce{^{210}Po}} emits $\alpha$ particles with an energy of 5.5\,MeV, that due to quenching will appear as a characteristic line of about 0.5\,MeV in the spectrum. For a long batch run (as assumed in Fig.~\ref{fig:c14po210}), the \ce{^{210}Po} decay rate could be determined at a 4\% precision. But even for a short day-long run (corresponding to continuous monitoring), 15\% precision could be reached for a cylindrical fiducial volume with 2\,m height and diameter. The minimum \ce{^{210}Po} abundance detectable within a day is $\sim 1\times10^{-24}$ g/g. Using $\alpha$/$\beta$ discrimination, this range could be further extended.

\paragraph{\ce{^{14}C}.} Already at an abundance of $10^{-18}$ relative to \ce{^{12}C} in the LS, \ce{^{14}C} will dominate the event rate in the very low energy-end of the spectrum. It reaches par with the expected external gamma background level at a few times $10^{-19}$. This is well below the JUNO requirement of $10^{-17}$. An exact rate measurement will less depend on the available event statistics but on the systematic understanding of the spectral shapes of $\gamma$ backgrounds and \ce{^{14}C} that are used in the fit.
\medskip\\
In conclusion, both backgrounds can be easily identified even if they are at or somewhat below the JUNO requirement levels. Precise measurements of the corresponding decay rates can be expected. However,  single backgrounds that are $\beta$-decays and do not feature distinctive spectral shapes are more difficult to determine. Studies will be resumed once experimental data becomes available.

\section{Conclusions}
The OSIRIS facility serves as the last stage in the purification chain of JUNO, measuring the residual radioactivity of the LS before being injected into the JUNO Central Detector. OSIRIS will provide valuable information on the efficiency of the purification chain and issue a timely alert in case problems occur with the radiopurity of the LS. In this article, we have given a detailed overview of the technical design of the OSIRIS detector. It has been optimized to screen the LS to the required purity levels of U/Th within days for {\it IBD-level} ($\leq$10$^{-15}$\,g/g) and 2--3 weeks for {\it solar-level} radiopurities ($\leq$10$^{-16\div17}$\,g/g). The final sensitivity and required measurement time depend on the level of radon emanation in the components of the purification chain and in liquid handling system. Further radio-isotopes that are expected to contaminate the LS are within reach of OSIRIS: excellent sensitivity will be achieved for the distinctive $^{14}$C and $^{210}$Po backgrounds.

The detector facility has the potential for a broader application once the LS filling of the JUNO Central Detector has been completed. Based on the existing detector geometry and the close-by LS purification system, OSIRIS can serve as test bench for the further development of liquid scintillators for JUNO and other applications. Both energy resolution and external gamma-ray shielding of the detector could be improved with relatively minor upgrades. These potentially open the path to test metal-loaded LS for a future $0\nu\beta\beta$ programme of JUNO or even a prevision measurement of the solar $pp$ neutrino flux within the OSIRIS setup. Corresponding studies are on-going.


\newpage 
\appendix 
\onecolumn
\section{Layout of the Liquid Scintillator Hall}
\label{app:LS_hall}
\begin{figure*}[h!]
    \centering
    \includegraphics[width=0.8\textwidth]{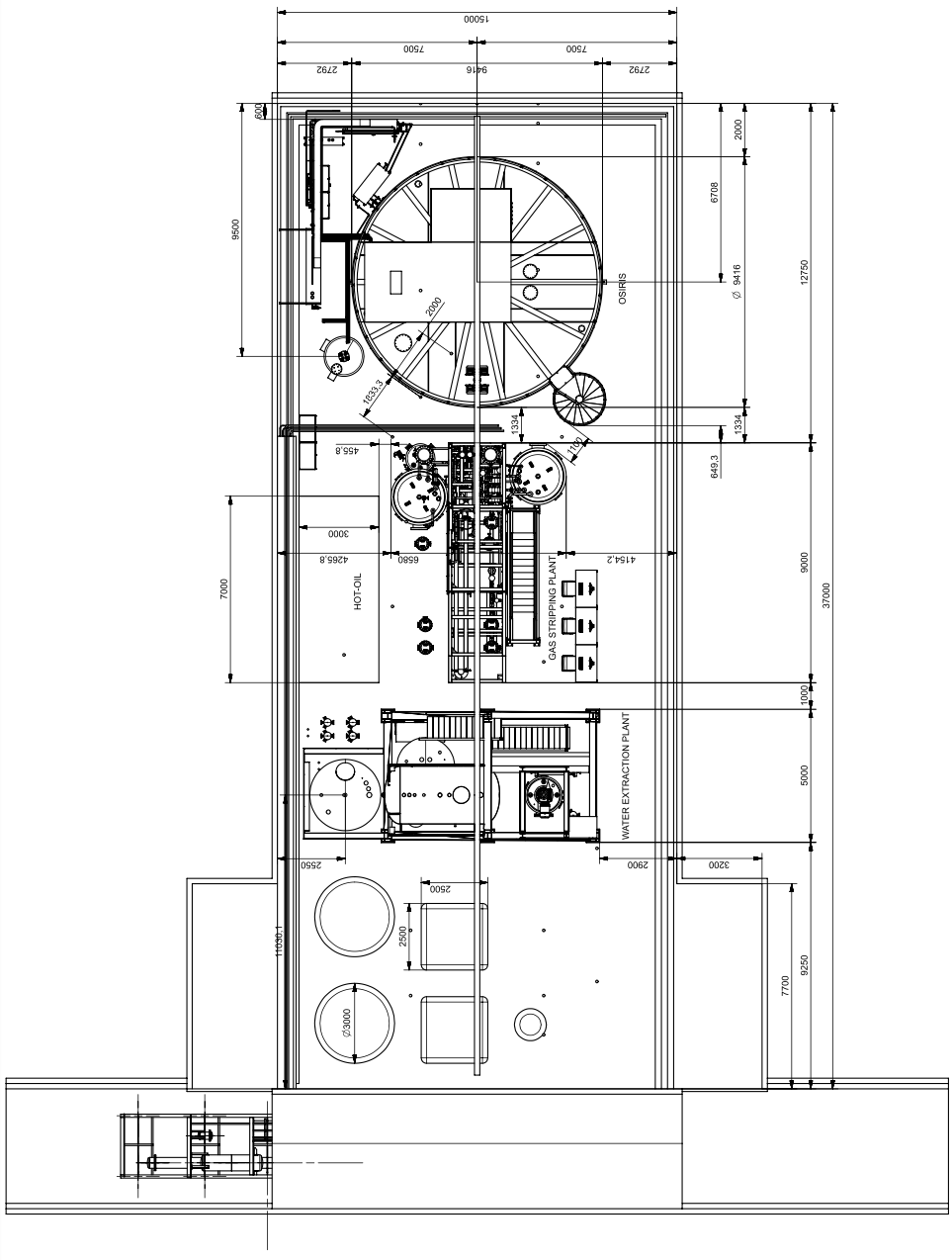}
    \caption{Layout of the underground Liquid Scintillator Hall. OSIRIS is located at the back of the hall, next to the Stripping Plant}
    \label{fig:LS_hall_layout}
\end{figure*}
\FloatBarrier
\begin{figure*}[h!]
    \centering
    \includegraphics[width=0.8\textwidth]{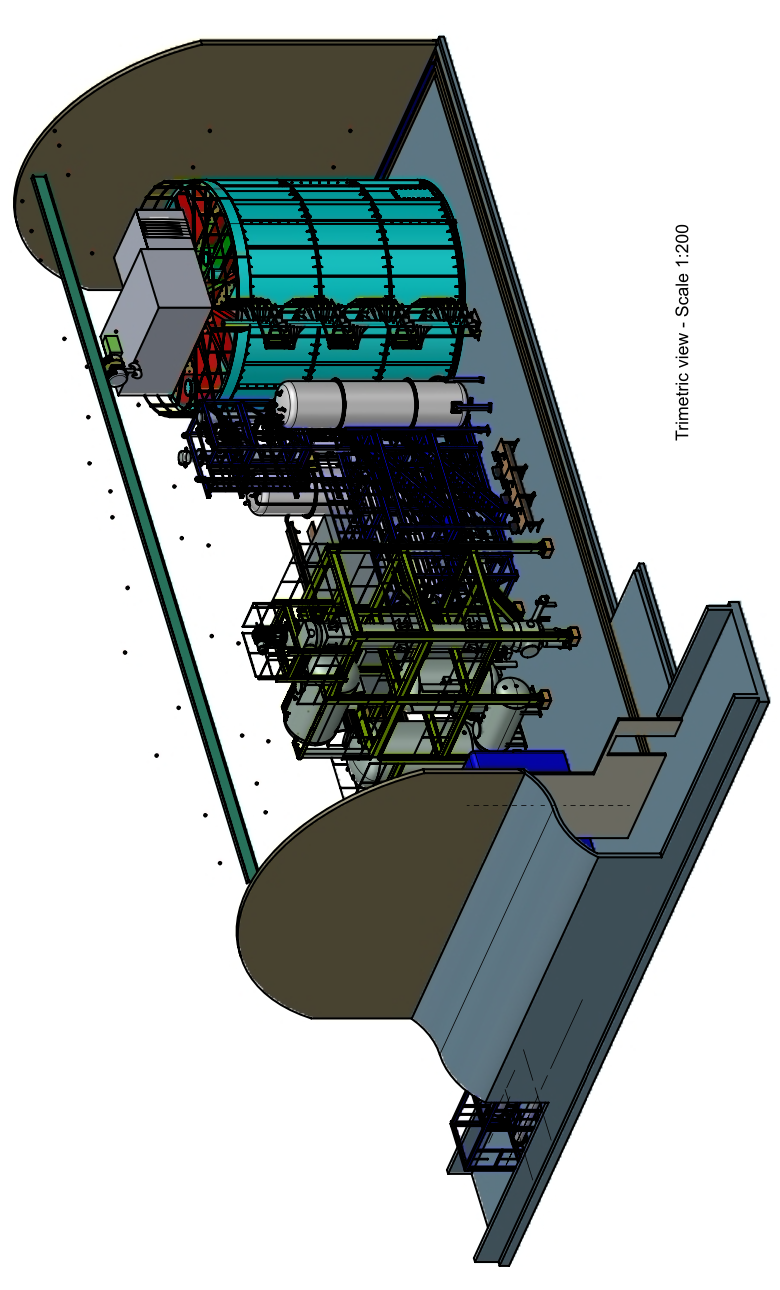}
    \caption{Isoview of the underground Liquid Scintillator Hall. OSIRIS is located at the back of the hall, next to the Stripping Plant}
    \label{fig:LS_hall_isoview}
\end{figure*}
\FloatBarrier
\newpage
\twocolumn
\bibliographystyle{spphys}       
\bibliography{main}   


\onecolumn
\begin{flushleft}
{\Large The JUNO Collaboration}\\
\bigskip
\input{juno_authors}\\
\bigskip
\input{juno_institutes}
\end{flushleft}

\end{document}

%% file: juno_authors.tex
Angel Abusleme$^{5}$, 
Thomas Adam$^{45}$, 
Shakeel Ahmad$^{66}$, 
Rizwan Ahmed$^{66}$, 
Sebastiano Aiello$^{55}$, 
Muhammad Akram$^{66}$, 
Fengpeng An$^{29}$, 
Guangpeng An$^{10}$, 
Qi An$^{22}$, 
Giuseppe Andronico$^{55}$
% ...

%% file: juno_institutes.tex
$^{1}$Yerevan Physics Institute, Yerevan, Armenia\\
$^{2}$Universit\'{e} Libre de Bruxelles, Brussels, Belgium\\
$^{3}$Universidade Estadual de Londrina, Londrina, Brazil\\
$^{4}$Pontificia Universidade Catolica do Rio de Janeiro, Rio, Brazil\\
$^{5}$Pontificia Universidad Cat\'{o}lica de Chile, Santiago, Chile\\
$^{6}$Universidad Tecnica Federico Santa Maria, Valparaiso, Chile\\
$^{7}$Beijing Institute of Spacecraft Environment Engineering, Beijing, China\\
$^{8}$Beijing Normal University, Beijing, China\\
$^{9}$China Institute of Atomic Energy, Beijing, China\\
$^{10}$Institute of High Energy Physics, Beijing, China\\
$^{11}$North China Electric Power University, Beijing, China\\
$^{12}$School of Physics, Peking University, Beijing, China\\
$^{13}$Tsinghua University, Beijing, China\\
$^{14}$University of Chinese Academy of Sciences, Beijing, China\\
$^{15}$Jilin University, Changchun, China\\
$^{16}$College of Electronic Science and Engineering, National University of Defense Technology, Changsha, China\\
$^{17}$Chongqing University, Chongqing, China\\
$^{18}$Dongguan University of Technology, Dongguan, China\\
$^{19}$Jinan University, Guangzhou, China\\
$^{20}$Sun Yat-Sen University, Guangzhou, China\\
$^{21}$Harbin Institute of Technology, Harbin, China\\
$^{22}$University of Science and Technology of China, Hefei, China\\
$^{23}$The Radiochemistry and Nuclear Chemistry Group in University of South China, Hengyang, China\\
$^{24}$Wuyi University, Jiangmen, China\\
$^{25}$Shandong University, Jinan, China, and Key Laboratory of Particle Physics and Particle Irradiation of Ministry of Education, Shandong University, Qingdao, China\\
$^{26}$Institute of Modern Physics, Chinese Academy of Sciences, Lanzhou, China\\
$^{27}$Nanjing University, Nanjing, China\\
$^{28}$Guangxi University, Nanning, China\\
$^{29}$East China University of Science and Technology, Shanghai, China\\
$^{30}$School of Physics and Astronomy, Shanghai Jiao Tong University, Shanghai, China\\
$^{31}$Tsung-Dao Lee Institute, Shanghai Jiao Tong University, Shanghai, China\\
$^{32}$Institute of Hydrogeology and Environmental Geology, Chinese Academy of Geological Sciences, Shijiazhuang, China\\
$^{33}$Nankai University, Tianjin, China\\
$^{34}$Wuhan University, Wuhan, China\\
$^{35}$Xi'an Jiaotong University, Xi'an, China\\
$^{36}$Xiamen University, Xiamen, China\\
$^{37}$School of Physics and Microelectronics, Zhengzhou University, Zhengzhou, China\\
$^{38}$Institute of Physics, National Yang Ming Chiao Tung University, Hsinchu\\
$^{39}$National United University, Miao-Li\\
$^{40}$Department of Physics, National Taiwan University, Taipei\\
$^{41}$Charles University, Faculty of Mathematics and Physics, Prague, Czech Republic\\
$^{42}$University of Jyvaskyla, Department of Physics, Jyvaskyla, Finland\\
$^{43}$IJCLab, Universit\'{e} Paris-Saclay, CNRS/IN2P3, 91405 Orsay, France\\
$^{44}$Univ. Bordeaux, CNRS, CENBG, UMR 5797, F-33170 Gradignan, France\\
$^{45}$IPHC, Universit\'{e} de Strasbourg, CNRS/IN2P3, F-67037 Strasbourg, France\\
$^{46}$Centre de Physique des Particules de Marseille, Marseille, France\\
$^{47}$SUBATECH, Universit\'{e} de Nantes,  IMT Atlantique, CNRS-IN2P3, Nantes, France\\
$^{48}$III. Physikalisches Institut B, RWTH Aachen University, Aachen, Germany\\
$^{49}$Institute of Experimental Physics, University of Hamburg, Hamburg, Germany\\
$^{50}$Forschungszentrum J\"{u}lich GmbH, Nuclear Physics Institute IKP-2, J\"{u}lich, Germany\\
$^{51}$Forschungszentrum J\"{u}lich GmbH, Central Institute of Engineering, Electronics and Analytics - Electronic Systems(ZEA-2), J\"{u}lich, Germany\\
$^{52}$Institute of Physics, Johannes-Gutenberg Universit\"{a}t Mainz, Mainz, Germany\\
$^{53}$Technische Universit\"{a}t M\"{u}nchen, M\"{u}nchen, Germany\\
$^{54}$Eberhard Karls Universit\"{a}t T\"{u}bingen, Physikalisches Institut, T\"{u}bingen, Germany\\
$^{55}$INFN Catania and Dipartimento di Fisica e Astronomia dell Universit\`{a} di Catania, Catania, Italy\\
$^{56}$Department of Physics and Earth Science, University of Ferrara and INFN Sezione di Ferrara, Ferrara, Italy\\
$^{57}$INFN Sezione di Milano and Dipartimento di Fisica dell Universit\`{a} di Milano, Milano, Italy\\
$^{58}$INFN Milano Bicocca and University of Milano Bicocca, Milano, Italy\\
$^{59}$INFN Milano Bicocca and Politecnico of Milano, Milano, Italy\\
$^{60}$INFN Sezione di Padova, Padova, Italy\\
$^{61}$Dipartimento di Fisica e Astronomia dell'Universit\`{a} di Padova and INFN Sezione di Padova, Padova, Italy\\
$^{62}$INFN Sezione di Perugia and Dipartimento di Chimica, Biologia e Biotecnologie dell'Universit\`{a} di Perugia, Perugia, Italy\\
$^{63}$Laboratori Nazionali di Frascati dell'INFN, Roma, Italy\\
$^{64}$University of Roma Tre and INFN Sezione Roma Tre, Roma, Italy\\
$^{65}$Institute of Electronics and Computer Science, Riga, Latvia\\
$^{66}$Pakistan Institute of Nuclear Science and Technology, Islamabad, Pakistan\\
$^{67}$Joint Institute for Nuclear Research, Dubna, Russia\\
$^{68}$Institute for Nuclear Research of the Russian Academy of Sciences, Moscow, Russia\\
$^{69}$Lomonosov Moscow State University, Moscow, Russia\\
$^{70}$Comenius University Bratislava, Faculty of Mathematics, Physics and Informatics, Bratislava, Slovakia\\
$^{71}$Department of Physics, Faculty of Science, Chulalongkorn University, Bangkok, Thailand\\
$^{72}$National Astronomical Research Institute of Thailand, Chiang Mai, Thailand\\
$^{73}$Suranaree University of Technology, Nakhon Ratchasima, Thailand\\
$^{74}$Department of Physics and Astronomy, University of California, Irvine, California, USA\\